\documentclass[aps,twocolumn,superscriptaddress,floatfix,longbibliography]{revtex4-2}
\usepackage{graphicx}
\usepackage{dcolumn}
\usepackage{bm}
\usepackage{subfigure}
\usepackage{amsmath}
\usepackage{mathrsfs}
\usepackage{amssymb}
\usepackage{setspace}
\usepackage{array}
\usepackage{multirow}
\usepackage{float}
\usepackage{flushend}
\usepackage{footmisc}

\usepackage{tikz}
\usepackage{graphicx}
\usetikzlibrary{calc}

\usepackage{xcolor}

\usepackage[pdfstartview=FitH,
CJKbookmarks=true,
colorlinks,
linkcolor=blue,
anchorcolor=blue,
citecolor=blue,
urlcolor=blue,
]{hyperref}

\usepackage{footmisc}

\renewcommand{\footnoterule}{
	\kern -4pt  
	\hrule width 0.18\linewidth height 0.6pt
	\kern 12pt 
}

\usepackage{footmisc}

\usepackage{braket}
\usepackage{braket}
\begin{document}

	\global\long\def\id{\mathbbm{1}}
	\global\long\def\ui{\mathbbm{i}}
	\global\long\def\ud{\mathrm{d}}

	\title{Symmetry-Filtered Relaxation Comb and Strong Quantum Mpemba Effect in Long-Range XXZ Spin Chains}
	\author{Zijun Wei}
	\affiliation{School of Physics, Nankai University, Tianjin 300071, China}
	
	\author{Mingdi Xu}
	\affiliation{School of Physics, Nankai University, Tianjin 300071, China}

    \author{Yefeng Song}
	\affiliation{School of Physics, Nankai University, Tianjin 300071, China}

    \author{Xiang-Ping Jiang}
	\affiliation{School of Physics, Hangzhou Normal University, Hangzhou 311121, China}

	\author{Yangqian Yan}
	\email{yqyan@cuhk.edu.hk}
	\affiliation{Department of Physics, The Chinese University of Hong Kong, Shatin, New Territories, Hong Kong, China}
	\affiliation{State Key Laboratory of Quantum Information Technologies and Materials, The Chinese University of Hong Kong, Hong Kong SAR, China}
	\affiliation{The Chinese University of Hong Kong Shenzhen Research Institute, 518057 Shenzhen, China}

	\author{Lei Pan}%
	\email{panlei@nankai.edu.cn}
	\affiliation{School of Physics, Nankai University, Tianjin 300071, China}

\begin{abstract}
We uncover a symmetry-filtered mechanism for anomalous dissipative relaxation in a long-range XXZ spin chain subject to local dephasing. At the isotropic point, the coherent Hamiltonian has global $SU(2)$ symmetry, whereas the full Liouvillian retains only the $U(1)$ symmetry associated with total magnetization. This structure pins a family of spatially uniform zero-$U(1)$-charge left eigenoperators with exact eigenvalues $\lambda=-2q$, forming a Liouvillian relaxation comb. For the ferromagnetic Dicke ground state, the overlap envelope on this comb is known exactly at finite size and becomes Gaussian in the large-$S$ limit. Since higher-$q$ components decay rapidly, the $q=1$ comb tooth controls the long-time dynamics and yields universal $D(t)\sim e^{-2t}$ relaxation independent of system size and interaction range. This mode-accessibility filtering realizes a spectral strong quantum Mpemba effect: an initially farther state relaxes faster than closer thermal states because slow non-steady Liouvillian modes are inaccessible. Weak breaking of the Hamiltonian $SU(2)$ symmetry restores slow-mode overlap and suppresses this acceleration.
\end{abstract}

	\maketitle

	\textit{\color{blue}Introduction.} Symmetry is a basic organizing principle of many-body physics, governing phases, conservation laws, and collective excitations. In open quantum systems, symmetries further constrain steady states, conserved quantities, and invariant subspaces of Lindblad dynamics~\cite{Lindblad1,Lindblad2,Breuer2002,Faz25,Alb13,Buc12}. How symmetry selects transient relaxation pathways in Liouville space, however, remains much less understood.

A striking manifestation of anomalous relaxation is the Mpemba effect, where an initially farther state relaxes faster than a closer one~\cite{ME,ME_classical1,ME_classical2,ME_classical3,ME_classical5,ME_classical9,Inverse_ME1,Inverse_ME2,Inverse_ME3,Tez25}. Recent theoretical and experimental developments have extended this phenomenon to quantum and dissipative settings, including open quantum systems, many-body dynamics, and non-Hermitian evolution~\cite{QME_Exp1,QME_Exp2,QME_Exp3,QME_Exp4,QME1,QME101,QME2,QME3,QME4,QME41,QME5,QME6,QME7,QME8,QME9,OpenQME1,OpenQME2,OpenQME3,OpenQME4,OpenQME5,OpenQME6,OpenQME7,OpenQME8,OpenQME10,OpenQME11,OpenQME12,OpenQME13,QME10,QME11,QME12,QME13,QME14,QME15,QME16,QME17,QME18,QME19,QME31,R1,R2,R3,Li26,Song26,Longhi26,Li25,Yamashika24,Yamashika25}. Despite these advances, a microscopic mechanism for robust strong quantum Mpemba effects~\cite{wei26,Kli17,Car21,Ryl23,Bao2026} in interacting many-body systems, beyond fine tuning or low-dimensional effective descriptions, is still needed.


In this Letter, we show that a robust strong quantum Mpemba effect can arise from symmetry-filtered Liouvillian mode accessibility in a long-range XXZ chain subject to local dephasing. The key point is not a change of the Liouvillian spectrum itself, but the selective accessibility of its modes from a high-symmetry initial state. At the isotropic point, the coherent Hamiltonian has global $SU(2)$ symmetry, while local dephasing reduces the manifest symmetry of the full Liouvillian to the $U(1)$ subgroup associated with total magnetization. This combination pins spatially uniform zero-$U(1)$-charge left eigenoperators with exact eigenvalues $\lambda=-2q$, forming a relaxation comb. The ferromagnetic Dicke ground state has nonzero overlap only with this comb; after the rapidly decaying higher-$q$ components disappear, the $q=1$ tooth produces universal $D(t)\sim e^{-2t}$ relaxation. Finite-temperature states, by contrast, retain overlap with slower Liouvillian sectors. As a result, the initially farther ground state can relax faster than closer thermal states, realizing a spectral strong quantum Mpemba effect. Breaking the Hamiltonian $SU(2)$ symmetry serves as a control: it restores slow-mode overlap and suppresses the acceleration.

\begin{figure*}[ht!]
		\centering
		\includegraphics[width=0.92\textwidth]{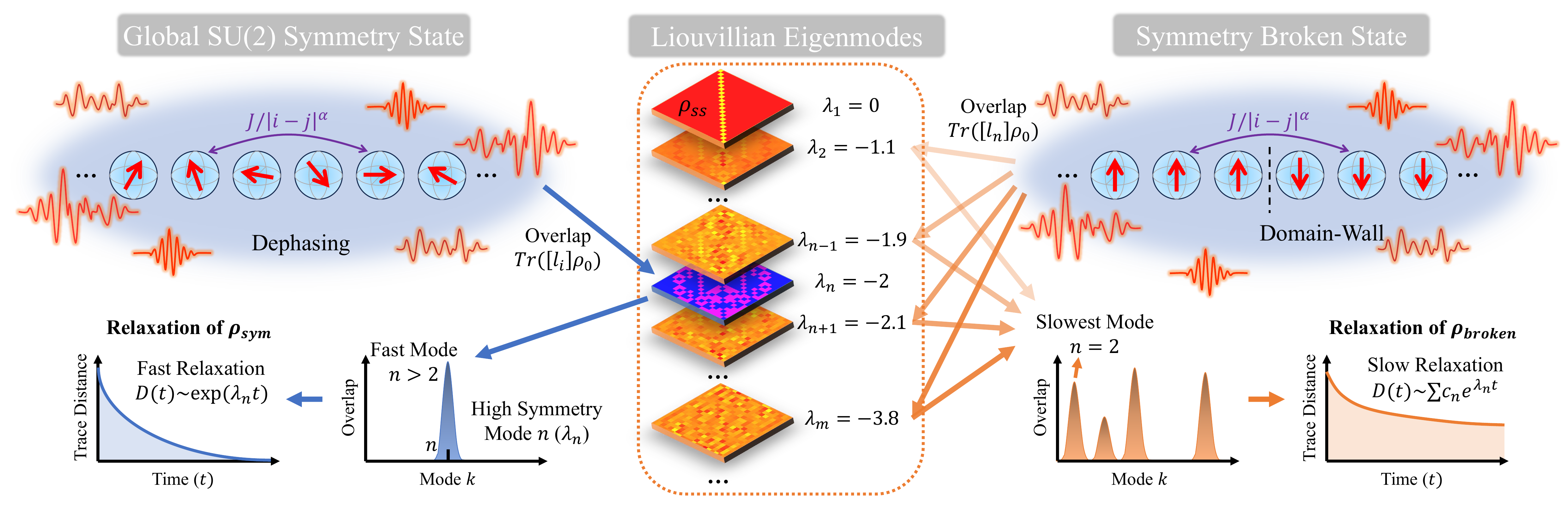}
		\caption{
			Schematic illustration of symmetry-filtered Liouvillian relaxation. 
			A highly symmetric initial state has nonzero overlap only with symmetry-compatible fast Liouvillian modes. 
			Although slower modes exist in the Liouvillian spectrum, they are inaccessible because their left eigenmodes belong to incompatible symmetry sectors. 
			The relaxation is therefore governed by the slowest accessible channel, producing an effectively single-channel exponential decay. 
			In contrast, a symmetry-broken initial state has finite overlap with multiple Liouvillian sectors, including slow modes, leading to slower relaxation and a hierarchy of dissipative timescales. 
			This symmetry-filtered mode accessibility underlies the strong quantum Mpemba effect discussed below.
		}
		\label{fig1}
	\end{figure*}

	\textit{\color{blue}Symmetry-Filtered Liouvillian Dynamics.}
	We consider an interacting open quantum spin system governed by the Lindblad master equation~\cite{Lindblad1,Lindblad2,Breuer2002}
	\begin{equation}
		\partial_t \rho
		=\mathcal{L}[\rho]
		=
		-i[H,\rho]
		+
		\sum_j
		\left(
		2L_j \rho L_j^\dagger
		-
		\{L_j^\dagger L_j,\rho\}
		\right),
		\label{eq:lindblad}
	\end{equation}
	where $H$ is the coherent many-body Hamiltonian and $L_j$ are dissipative jump operators. The Liouvillian superoperator \(\mathcal{L}\) governs the full nonequilibrium dynamics in Liouville space. The relaxation dynamics can be resolved in terms of Liouvillian eigenmodes,
	\begin{equation}
		\rho(t)=\rho_{\rm ss}+\sum_{k=2}^{d^2} c_k e^{\lambda_k t} r_k,
		\label{eq:liouvillian_decomposition}
	\end{equation}
	where \(\rho_{\rm ss}=\lim_{t\rightarrow\infty}\rho(t)\) is the steady state, \(\lambda_k\) and \(r_k\) are Liouvillian eigenvalues and right eigenmodes, namely, \(\mathcal{L}[r_k]=\lambda_k r_k\). The eigenvalues are sorted by their real parts such that $0 = \text{Re}(\lambda_1) \geq \text{Re}(\lambda_2) \geq \dots \geq \text{Re}(\lambda_{d^2})$ with Hilbert space dimension $d$, and \(c_k\equiv \text{Tr}(l_k^\dagger\rho_0)\) are overlap coefficients determined by the initial state with the corresponding left eigenmodes \(l_k\) satisfying \(\mathcal{L}^\dagger [l_k]=\lambda^*_k l_k\). 
	
	Equation~\eqref{eq:liouvillian_decomposition} shows that the Liouvillian gap alone does not necessarily determine the observed relaxation rate. A mode contributes to the dynamics only if its overlap coefficient is nonzero. Therefore, for a given initial state, the relevant asymptotic decay rate is set by the slowest \emph{accessible} Liouvillian mode,
	\begin{equation}
		\Gamma_{\rm eff}(\rho_0)
		=
		-\max_{n>1:\,c_n\neq 0}\mathrm{Re}\,\lambda_n .
		\label{eq:effective_gap}
	\end{equation}
	Here \(\Gamma_{\rm eff}\) is an initial-state-dependent effective relaxation rate. It can be larger than the Liouvillian gap if conventional slow non-steady modes have vanishing overlap with the chosen initial state.
This observation provides the basis for symmetry-filtered relaxation. If symmetry decomposes Liouville space into dynamically disconnected sectors~\cite{Alb13,Buc12}, an initial state can relax only through sectors to which it has nonzero projection. Thus relaxation is determined not only by the spectrum of \(\mathcal{L}\), but also by the symmetry-constrained accessibility of its eigenmodes. A high-symmetry state may bypass conventional slow modes and relax through a faster accessible channel, as illustrated in Fig.~\ref{fig1}.

\begin{figure}[ht!]
		\centering
		\includegraphics[width=0.98\linewidth]{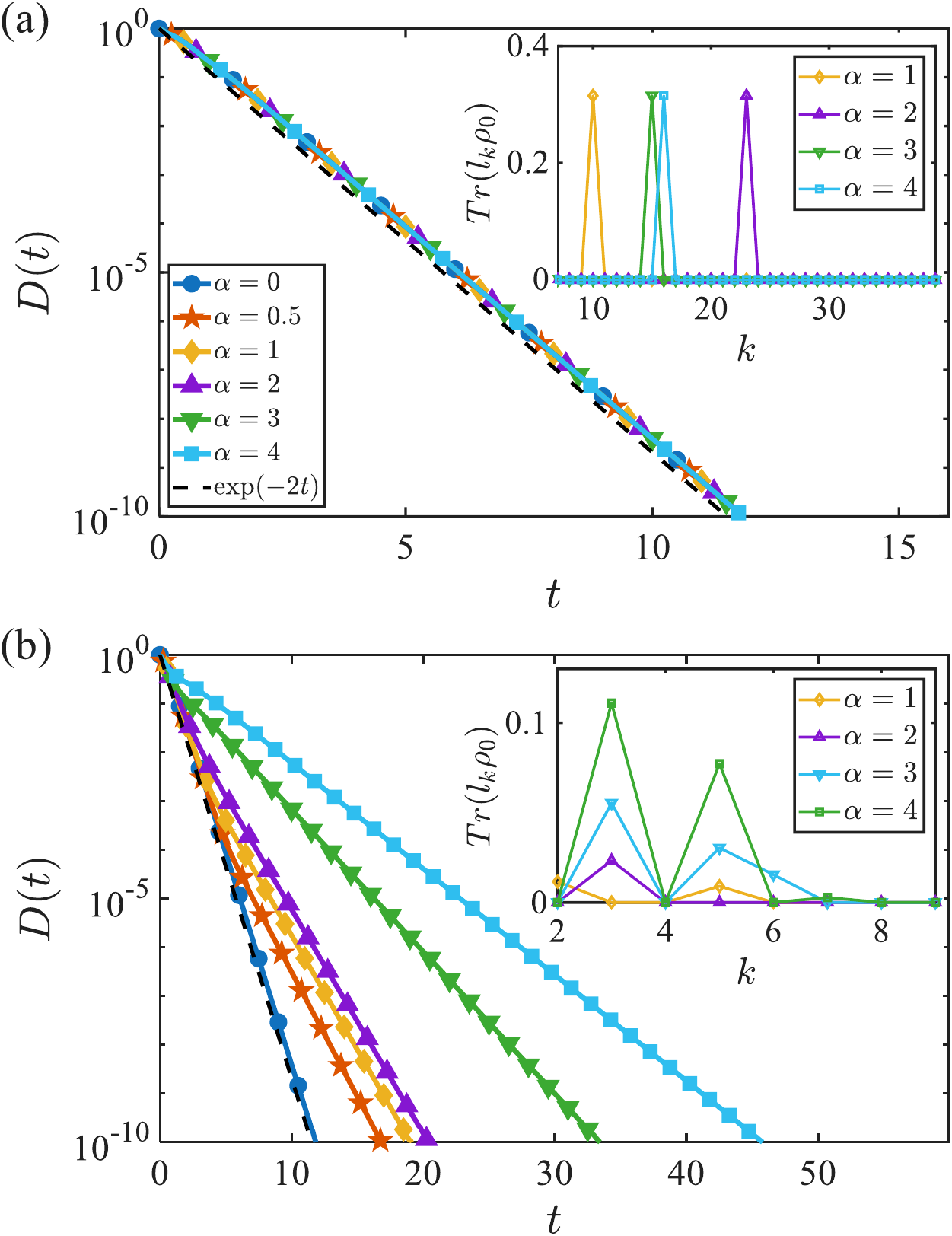}
		\caption{
			Ground-state relaxation dynamics in the long-range XXZ model (\(L=10\), ferromagnetic exchange \(J=1\) in Eq.~\eqref{ham}).
			(a) In the \(SU(2)\)-symmetric XXX limit (\(\Delta=1\)), the ground-state dynamics are supported only on the exact \(\lambda=-2q\) Liouvillian modes. Higher-order components decay as \(e^{-2qt}\) with \(q>1\), leaving the long-time relaxation universally governed by \(D(t)\sim e^{-2t}\). The inset shows that the asymptotic spectral weight is confined to the \(\lambda=-2\) mode, demonstrating symmetry-selected access to an isolated fast-decay channel.
			(b) Breaking the symmetry through anisotropy (\(\Delta=1.1\)) fragments the Liouvillian spectrum and activates multiple decay sectors, including slow modes. The resulting slow-mode overlap suppresses the relaxation dynamics and produces a pronounced interaction-range-dependent slowdown.
		}
		\label{fig2}
	\end{figure}

	\textit{\color{blue}Symmetry-Filtered Fast Relaxation in Long-Range Spin Chains.} We consider the dissipative dynamics of the long-range spin-$1/2$ XXZ chain~\cite{J1J2_1,J1J2_2,XXZ_exp1,XXZ_exp2,XXZ_exp3,XXZ_exp4,XXZ_exp5,Def23,Ric14,Jur14,Hau13b,Jos21,Yamashika26}
	\begin{equation}
		H=
		\sum_{i<j}\frac{-J}{|i-j|^\alpha}
		\left(
		S_i^xS_j^x
		+
		S_i^yS_j^y
		+
		\Delta S_i^zS_j^z
		\right),
		\label{ham}
	\end{equation}
	where the interaction strength \(J>0\), and the minus sign in Eq.~\eqref{ham} denotes ferromagnetic exchange. Here \(\alpha\) controls the interaction range, with \(\alpha\to\infty\) corresponding to the short-range limit and \(\alpha=0\) to all-to-all interactions. At the isotropic point \(\Delta=1\), the ferromagnetic ground-state manifold lies in the maximum-spin sector \(S=L/2\). We work with its symmetry-unbroken Dicke representative within a fixed magnetization sector.
	The chain is subject to local dephasing dissipation described by the Lindblad equation \eqref{eq:lindblad} with jump operators
	\begin{equation}
		L_j=\frac{\sigma_j^z+1}{2}.
	\end{equation}
	We focus primarily on the isotropic point \(\Delta=1\), where the Hamiltonian possesses global \(SU(2)\) symmetry. The full Liouvillian, however, does not retain this \(SU(2)\) symmetry because the dephasing jump operators single out the \(z\) direction. Instead, the Lindblad dynamics preserves the \(U(1)\) symmetry associated with total magnetization \(S^z\).
	To characterize relaxation, we compute the trace distance between the evolving density matrix \(\rho(t)\) and the steady state \(\rho_{\mathrm{ss}}\)
	\begin{equation}
		D(t)=\frac{1}{2}\|\rho(t)-\rho_{\mathrm{ss}}\|_1,
		\label{distance}
	\end{equation}
	where $\|O\|_1\equiv\text{Tr}(\sqrt{O^\dagger O})$.
    For pure dephasing dynamics, \(\rho_{\mathrm{ss}}\) is the infinite-temperature maximally mixed state within the corresponding magnetization sector.
	
	We first consider the zero-temperature ground state at the \(SU(2)\)-symmetric point, namely the symmetry-unbroken ferromagnetic ground-state representative in the chosen fixed-\(S^z\) sector. Remarkably, independent of both system size and interaction exponent \(\alpha\), the relaxation exhibits the universal exponential form
	\begin{equation}
		D(t)\sim e^{-2t},
		\label{universal}
	\end{equation}
	as shown in Fig.~\ref{fig2}(a). This universality persists throughout the long-range regime \(\alpha>0\), despite substantial changes in the microscopic many-body spectrum induced by varying the interaction range.
	
	To uncover the origin of this behavior, we use the Liouvillian-mode expansion introduced in Eq.~\eqref{eq:liouvillian_decomposition}. Although the long-time dynamics of generic states are usually controlled by the slowest nonzero Liouvillian modes, the ground state avoids these slow sectors entirely because its spectral weight is symmetry-filtered.
	The inset of Fig.~\ref{fig2}(a) resolves the overlap between the initial ground state and the Liouvillian eigenmodes. Although the dominant mode index changes with \(\alpha\), the nonvanishing spectral weight is always confined to modes with eigenvalue $\lambda=-2q$ (with only the slowest evolving $\lambda=-2$ mode displayed). Thus the relaxation bypasses conventional slow modes and proceeds exclusively through a symmetry-selected fast-decay channel.
	
	This selection rule originates from the global $SU(2)$ symmetry of the Hamiltonian, together with the $U(1)$ charge conservation retained by the full Liouvillian. At \(\Delta=1\), the Hamiltonian commutes with the total-spin generators $	S^\pm=\sum_i\sigma_i^\pm,
	\qquad
	S^z=\frac12\sum_i\sigma_i^z$. The essential observation is that the high-symmetry ground state has strongly constrained spectral support in Liouville space. Its nonvanishing overlap is restricted to a symmetry-selected family of spatially uniform zero-\(U(1)\)-charge modes, while spatially nonuniform slow modes are filtered out by symmetry mismatch and do not participate in the relaxation. The selected relaxation channels are governed by a family of spatially uniform operators in the zero-\(U(1)\)-charge sector,
	\begin{equation}
		O^{(2q)}
		=
		\sum_{\substack{i_1,\ldots,i_q,j_1,\ldots,j_q\\ {\rm all\ distinct}}}
		\sigma_{i_1}^+\cdots\sigma_{i_q}^+
		\sigma_{j_1}^-\cdots\sigma_{j_q}^- ,
		\label{eq:O2q}
	\end{equation}
	where \(q\in\mathbb{Z}^+\). These operators contain equal numbers of raising and lowering operators and therefore commute with \(S^z\), i.e., they lie in the zero-\(U(1)\)-charge sector of the full Liouvillian.
	
	The $SU(2)$ symmetry of the coherent Hamiltonian, rather than of the full Liouvillian, prevents these spatially uniform collective operators from being mixed by the Hamiltonian evolution into nonuniform operator sectors. For the representative case \(q=1\),
	\[
	O^{(2)}=\sum_{i\neq j}\sigma_i^+\sigma_j^-
	=
	S^+S^- - S^z - \frac{L}{2}.
	\]
	Since \([H,S^\pm]=[H,S^z]=0\) at \(\Delta=1\), one obtains \([H,O^{(2)}]=0\). Thus the coherent part of the adjoint Liouvillian does not mix \(O^{(2)}\) with other operator sectors. The local dephasing dissipator acts on each transverse spin operator with a fixed decay rate, giving
	\[
	\mathcal{L}^\dagger(O^{(2)})=-2O^{(2)}.
	\]
	More generally, the spatially uniform operators \(O^{(2q)}\) are permutation-invariant and can be expressed in the collective-spin algebra generated by \(S^\pm\) and \(S^z\). Since the isotropic Hamiltonian commutes with all collective spin generators, \([H,O^{(2q)}]=0\) for all \(q\), the local dephasing dissipator then counts the number \(2q\) of transverse spin-flip operators, yielding
	\begin{equation}
		\mathcal{L}^\dagger\!\left(O^{(2q)}\right)
		=
		-2q\,O^{(2q)} .
		\label{eq:O2q_decay}
	\end{equation}
	Thus the \(\lambda=-2q\) modes form an exact family of spatially uniform Liouvillian left eigenoperators in the zero-\(U(1)\)-charge sector whose eigenvalues are independent of system size and interaction range.

	For the symmetry-unbroken ferromagnetic ground state, the overlap with this relaxation comb has an exact finite-size form. In the zero-magnetization sector and for \(L=2S\), the ground-state representative is the Dicke state \(|S=L/2,M=0\rangle\). After normalizing the \(2q\)-spin coherence shell, the overlap amplitudes and spectral weights are
	\begin{equation}
		c_q=
		\frac{\binom{S}{q}}{\sqrt{\binom{2S}{S}}},
		\qquad
		w_q=|c_q|^2=
		\frac{\binom{S}{q}^2}{\binom{2S}{S}},
		\qquad
		q=0,1,\ldots,S .
		\label{eq:comb_weight_main}
	\end{equation}

The zero-magnetization expression is a special case of a more general fixed-magnetization formula. For a Dicke representative \(|S,M\rangle\), with \(S=L/2\), the numbers of up and down spins are \(N_\uparrow=S+M\) and \(N_\downarrow=S-M\), respectively. The normalized overlap amplitude with the \(2q\)-spin coherence shell is
\begin{equation}
	c_q(M)
	=
	\sqrt{
	\frac{
	\binom{S+M}{q}\binom{S-M}{q}
	}{
	\binom{2S}{S+M}
	}
	},
	\qquad
	q=0,1,\ldots,S-|M| .
	\label{eq:comb_overlap_M_main}
\end{equation}
The corresponding spectral weight is \(w_q(M)=|c_q(M)|^2\).
For \(M=0\), Eq.~\eqref{eq:comb_overlap_M_main} reduces to
Eq.~\eqref{eq:comb_weight_main}. The detailed derivation and
large-\(S\) limit are given in the Supplemental Material.
  For the zero-magnetization sector, the large-\(S\) limit of \(w_q\) approaches a Gaussian envelope centered at \(q=S/2=L/4\),
\begin{equation}
	w_q
	\simeq
	\frac{1}{\sqrt{2\pi\sigma_q^2}}
	\exp\!\left[
	-\frac{(q-L/4)^2}{2\sigma_q^2}
	\right],
	\qquad
	\sigma_q^2\simeq \frac{L}{16}.
	\label{eq:comb_gaussian_main}
\end{equation}
Thus the finite-size Liouvillian comb is accompanied by an
analytically controlled thermodynamic-limit envelope.

The high-symmetry ground-state density matrix has nonzero spectral weight only on this symmetry-selected family, while its overlap with conventional slow modes vanishes. Higher-order modes with \(q>1\) decay as \(e^{-2qt}\) and quickly disappear, leaving the slowest accessible member \(O^{(2)}\) to dominate the long-time dynamics. This produces the universal \(D(t)\sim e^{-2t}\) relaxation, which we verify up to \(L=14\) in the Supplemental Material.
The role of symmetry can be directly tested by introducing anisotropy. As shown in Fig.~\ref{fig2}(b), at \(\Delta=1.1\) the Liouvillian spectrum fragments, the initial state acquires overlap with multiple decay sectors including slow modes, and the relaxation becomes interaction-range dependent. Thus anomalous fast relaxation originates not from long-range interactions alone, but from symmetry-filtered Liouvillian mode accessibility. Weak breaking of the Hamiltonian \(SU(2)\) symmetry only perturbatively activates the slow tail, as shown in the Supplemental Material.

\begin{figure}[t]
		\centering
		\includegraphics[width=0.98\columnwidth]{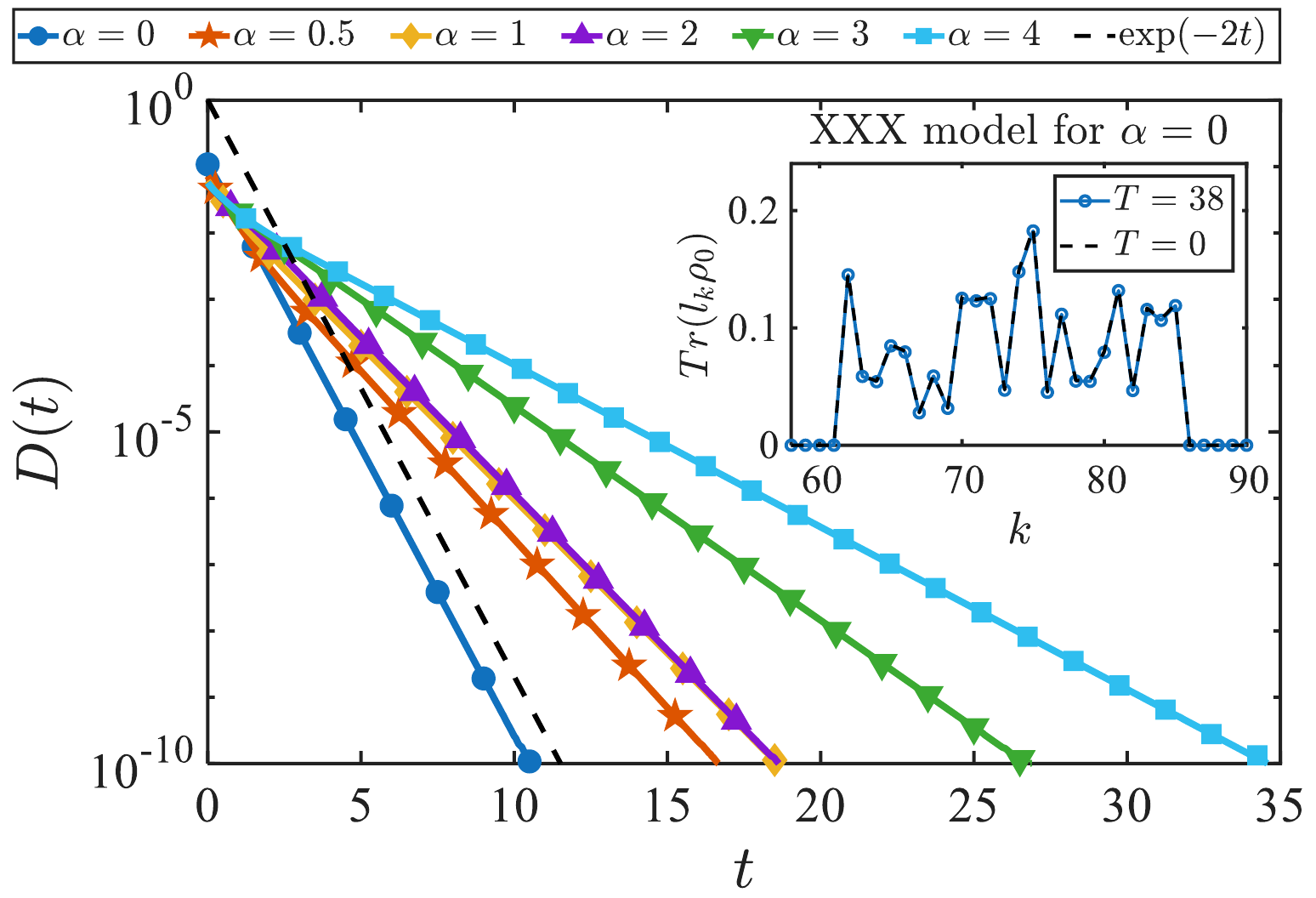}
		
\caption{Emergence and disappearance of the strong quantum Mpemba effect. The dashed line denotes the universal symmetry-filtered trajectory \(D(t)\sim e^{-2t}\) of the symmetry-unbroken ferromagnetic ground state. Solid curves represent finite-temperature initial states for different interaction exponents \(\alpha\). For the finite $\alpha>0$ values shown, the ground state is initially farther from the infinite-temperature steady state but relaxes faster than the thermal states and eventually overtakes them, demonstrating a strong quantum Mpemba effect. In the all-to-all limit \((\alpha=0)\), the enlarged permutation symmetry produces a degenerate Liouvillian manifold at \(\lambda=-2\). As a result, the spectral separation between the ground-state and thermal relaxation channels is lost, the dominant asymptotic decay rates become
indistinguishable, and the strong-Mpemba separation disappears. Inset: spectral overlap at \(\alpha=0\), showing that both the \(T=0\) and finite-temperature states have support within the same degenerate \(\lambda=-2\) manifold.}
\label{fig3}
\end{figure}

	\begin{figure*}[t]
		\centering
		\includegraphics[width=0.90\textwidth]{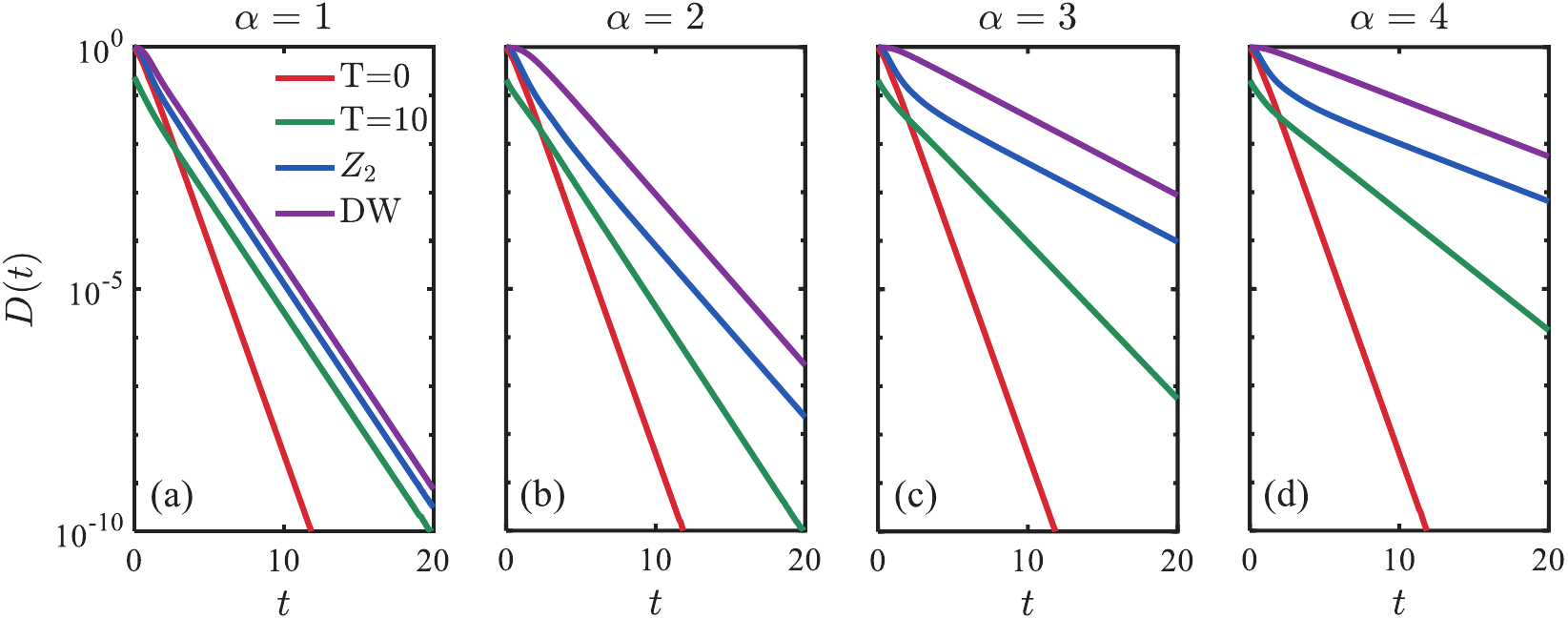}
		\caption{
			Interaction-range evolution and symmetry-breaking control of relaxation. 
			The relaxation dynamics of the ground state (\(T=0\)), thermal state (\(T=10\)), $Z_2$-symmetric state, and domain-wall state are shown for different interaction exponents: (a) \(\alpha=1\), (b) \(\alpha=2\), (c) \(\alpha=3\), and (d) \(\alpha=4\). The highly symmetric ground state remains dominated by the isolated fast-decay Liouvillian channel at \(\lambda=-2\), whereas lower-symmetry states develop finite overlap with slow modes and therefore relax more slowly. The interaction range controls the separation between relaxation trajectories, while the hierarchy is fixed by symmetry-filtered Liouvillian accessibility.
		}
		\label{fig4}
	\end{figure*}

	\textit{\color{blue}Strong Quantum Mpemba Effect.} We compare the ground-state dynamics with finite-temperature initial states,
\begin{equation}
	\rho_T
	=
	\frac{1}{Z}
	\sum_n
	e^{-\beta E_n}
	|\psi_n\rangle\langle\psi_n|,
\end{equation}
where \(\beta=1/T\). Figure~\ref{fig3} shows that, for the finite interaction ranges studied here, the symmetry-unbroken ground state follows $D(t)\sim e^{-2t}$, whereas thermal states are initially closer to $\rho_{\rm ss}$ but relax more slowly. This is a strong quantum Mpemba effect in the spectral sense: the ground state has zero overlap with slow non-steady modes and hence \(\Gamma_{\rm gs}=2\), while the thermal states shown in Fig.~\ref{fig3} retain finite slow-mode overlap and have \(\Gamma_T<2\). Equivalently, the high-symmetry state couples only to the symmetry-selected \(\lambda=-2q\) comb, with the exact overlap envelope in Eq.~\eqref{eq:comb_weight_main}, whereas finite-temperature states access slower Liouvillian sectors. The relaxation speed is therefore controlled by mode accessibility, not only by the initial distance to the steady state.

The all-to-all limit \(\alpha=0\) is qualitatively different. The isotropic ferromagnetic Hamiltonian becomes fully permutation symmetric,
\begin{equation}
	H_{\alpha=0}
	=
	-\frac{J}{L}\sum_{i<j}\mathbf{S}_i\cdot\mathbf{S}_j
	=
	-\frac{J}{2L}
	\left(
	\mathbf{S}_{\rm tot}^2-\frac{3L}{4}
	\right),
	\label{eq:all_to_all_H}
\end{equation}
which produces a highly degenerate Liouvillian manifold at \(\lambda=-2\). As a result, the spectral separation between ground-state and thermal relaxation channels for \(\alpha>0\) is lost; the dominant asymptotic rates become indistinguishable and the strong-Mpemba separation disappears, as confirmed by the inset of Fig.~\ref{fig3}.

\textit{\color{blue}Interaction-Range Evolution and Symmetry Control.} The same mechanism provides a way to organize dissipative timescales by the symmetry structure of the initial state. Figure~\ref{fig4} compares the ground state, a finite-temperature thermal state, a \(Z_2\)-symmetric state, and a domain-wall state for different interaction exponents \(\alpha\). A clear hierarchy emerges: the high-symmetry ground state remains dominated by the isolated \(\lambda=-2\) channel, whereas lower-symmetry states acquire overlap with slow modes and relax more slowly.

This hierarchy persists across the long-range interacting regime. Varying \(\alpha\) changes the microscopic spectrum and the quantitative separation between trajectories, but the ordering of relaxation times is controlled primarily by which Liouvillian sectors are accessible from the initial state. Symmetry breaking can therefore serve as a practical knob for tuning dissipative lifetimes.

\textit{\color{blue}Conclusion.---}
We have shown that dissipative relaxation in an interacting long-range XXZ chain can be organized by a symmetry-filtered Liouvillian comb. At the isotropic point, although the full Liouvillian retains only the $U(1)$ symmetry associated with total magnetization, the Hamiltonian $SU(2)$ symmetry pins a family of spatially uniform zero-$U(1)$-charge left eigenoperators with eigenvalues $\lambda=-2q$. For the symmetry-unbroken ferromagnetic ground state, the overlap envelope on this comb is known exactly at finite size and approaches a Gaussian form in the large-\(S\) limit. Although the initial spectral weight is distributed over many comb teeth, higher-\(q\) components decay rapidly as \(e^{-2qt}\), leaving the \(q=1\) tooth to produce universal \(D(t)\sim e^{-2t}\) relaxation independent of system size and interaction range.

This mechanism demonstrates that relaxation is governed not only by the Liouvillian spectrum or gap, but also by the symmetry-constrained accessibility of its modes. In the symmetric regime, the high-symmetry ground state has vanishing overlap with conventional slow non-steady modes, whereas breaking the Hamiltonian \(SU(2)\) symmetry restores slow-mode accessibility and suppresses relaxation. This contrast yields a spectral strong quantum Mpemba effect: an initially farther high-symmetry state relaxes faster because slow modes are filtered out, while closer thermal states retain finite slow-mode overlap. Our results identify symmetry-filtered Liouvillian mode accessibility as a general organizing principle for anomalous relaxation and suggest a route to engineering dissipative timescales in long-range quantum simulators.

\textit{\color{blue}Acknowledgments.} We thank Zhendong Zhang for helpful discussions. The work is supported by the National Natural Science Foundation of China (Grant No. 12304290, No. 12505017), Beijing National Laboratory for Condensed Matter Physics (2025BNLCMPKF017), and the Fundamental Research Funds for the Central Universities. \\

\bibliography{References}

\clearpage
\onecolumngrid
\section*{Supplemental Material}

 In the Supplemental Material, we provide additional details on the Liouvillian spectral decomposition and the analytical derivation of the exact $\lambda=-2q$ left-Liouvillian eigenoperator family. We also analyze the finite-size stability of the universal comb relaxation and the robustness of the dynamics against weak breaking of the Hamiltonian $SU(2)$ symmetry.
We further derive the exact finite-size expression for the overlap between the ferromagnetic Dicke ground state and the symmetry-selected zero-$U(1)$-charge $\lambda=-2q$ left Liouvillian eigenoperators, together with its large-$S$ form in the thermodynamic limit. Additional results include the all-to-all degeneracy analysis and the extension of the comb-overlap structure to nonzero magnetization sectors.

\section*{Spectral Decomposition of the Liouvillian Superoperator}
\label{App1}

We first summarize the spectral decomposition used in the main text in
operator-space notation. The Lindblad generator is written as
\begin{equation}
	\partial_t \rho
	=
	\mathcal L[\rho]
	=
	-i[H,\rho]
	+
	\sum_j
	\left(
	2L_j\rho L_j^\dagger
	-
	\{L_j^\dagger L_j,\rho\}
	\right),
	\label{Aeq1}
\end{equation}
where $H$ is the coherent Hamiltonian and $L_j$ are the jump operators.
The Liouvillian $\mathcal L$ is a linear superoperator acting on the
$d^2$-dimensional operator space, where $d$ is the Hilbert-space dimension
within the symmetry sector under consideration.

The right Liouvillian eigenoperators $r_n$ and the corresponding left
eigenoperators $l_n$ are defined by
\begin{equation}
	\mathcal L[r_n]
	=
	\lambda_n r_n,
	\qquad
	\mathcal L^\dagger[l_n]
	=
	\lambda_n^* l_n ,
	\label{Aeq2}
\end{equation}
where $\mathcal L^\dagger$ denotes the adjoint superoperator with respect
to the Hilbert--Schmidt inner product,
\begin{equation}
	\langle A,B\rangle_{\rm HS}
	=
	\mathrm{Tr}(A^\dagger B).
\end{equation}
We choose the biorthogonal normalization
\begin{equation}
	\mathrm{Tr}(l_m^\dagger r_n)
	=
	\delta_{mn}.
	\label{Aeq3}
\end{equation}
The eigenvalues are ordered according to their real parts,
\begin{equation}
	0=\mathrm{Re}\,\lambda_1
	\geq
	\mathrm{Re}\,\lambda_2
	\geq
	\cdots ,
\end{equation}
with $r_1=\rho_{\rm ss}$ denoting the steady state.

With this normalization, the deviation of the initial state from the steady
state can be expanded as
\begin{equation}
	\rho(0)-\rho_{\rm ss}
	=
	\sum_{n>1}
	c_n r_n,
	\qquad
	c_n
	=
	\mathrm{Tr}
	\left[
	l_n^\dagger
	\left(\rho(0)-\rho_{\rm ss}\right)
	\right].
	\label{Aeq4}
\end{equation}
For transient modes $n>1$, the biorthogonality condition implies
$\mathrm{Tr}(l_n^\dagger\rho_{\rm ss})=0$, so the coefficient can be written
equivalently as
\begin{equation}
	c_n
	=
	\mathrm{Tr}
	\left(
	l_n^\dagger \rho(0)
	\right).
	\label{Aeq5}
\end{equation}
The time evolution then takes the form
\begin{equation}
	\rho(t)-\rho_{\rm ss}
	=
	\sum_{n>1}
	c_n e^{\lambda_n t} r_n .
	\label{Aeq6}
\end{equation}
This is the Liouvillian-mode expansion used throughout the main text.

Equation~\eqref{Aeq6} shows that the observed relaxation rate for a given
initial state is not necessarily determined by the Liouvillian gap alone.
A mode contributes to the dynamics only if its overlap coefficient $c_n$ is
nonzero. Therefore, the relevant asymptotic decay rate is the slowest
accessible rate,
\begin{equation}
	\Gamma_{\rm eff}(\rho_0)
	=
	-\max_{n>1:\,c_n\neq0}
	\mathrm{Re}\,\lambda_n .
	\label{Aeq7}
\end{equation}
If the overlap with the conventional slow non-steady modes vanishes, the
effective relaxation rate can be larger than the Liouvillian gap. This is
the spectral basis of the symmetry-filtered relaxation mechanism. In the
present problem, the ferromagnetic Dicke state has nonzero overlap only with
the symmetry-selected zero-$U(1)$-charge comb modes, so its late-time
relaxation is governed by the slowest accessible comb tooth at
$\lambda=-2$.

For numerical diagonalization, we use a matrix representation of the same
superoperator obtained by vectorizing the density matrix through the
Choi--Jamiolkowski/operator-state mapping~\cite{Choi1975,Jamiolkowski1972,Prosen2008,Talkington2024}.
Denoting the vectorized density matrix by $|\rho\rangle\rangle$, the
Liouvillian action is written as
\begin{equation}
	|\mathcal L[\rho]\rangle\rangle
	=
	\mathbb L |\rho\rangle\rangle ,
\end{equation}
where
\begin{equation}
	\mathbb L
	=
	-i
	\left(
	H\otimes I
	-
	I\otimes H^{\mathrm T}
	\right)
	+
	\sum_j
	\left[
	2L_j\otimes L_j^*
	-
	L_j^\dagger L_j\otimes I
	-
	I\otimes
	(L_j^\dagger L_j)^{\mathrm T}
	\right].
	\label{Aeq8}
\end{equation}
The Liouvillian eigenvalues and eigenmodes shown in the figures are obtained
by direct diagonalization of $\mathbb L$ in the relevant fixed-magnetization
sector, and then mapped back to operator space to compute the overlaps
$c_n=\mathrm{Tr}(l_n^\dagger\rho_0)$.

\section*{Independent mode decay and channel decoupling} \label{App2}
Equation~\eqref{Aeq6} shows that Liouvillian relaxation is a linear superposition of independently decaying eigenmodes. The time-dependent coefficient of each mode is
\begin{equation}\label{Aeq4b}
	c_k(t)=c_k(0)e^{\lambda_k t} .
\end{equation}
Thus, if \(c_k(0)=0\) for a given mode, this coefficient remains zero for all times. There is no dynamical transfer of spectral weight between different Liouvillian eigenmodes. This simple property is central to the symmetry-filtering mechanism: once a symmetry forbids the overlap between an initial state and a slow Liouvillian mode, that slow channel remains inaccessible throughout the evolution.

\begin{figure}
	\centering
	\includegraphics[width=0.98\linewidth]{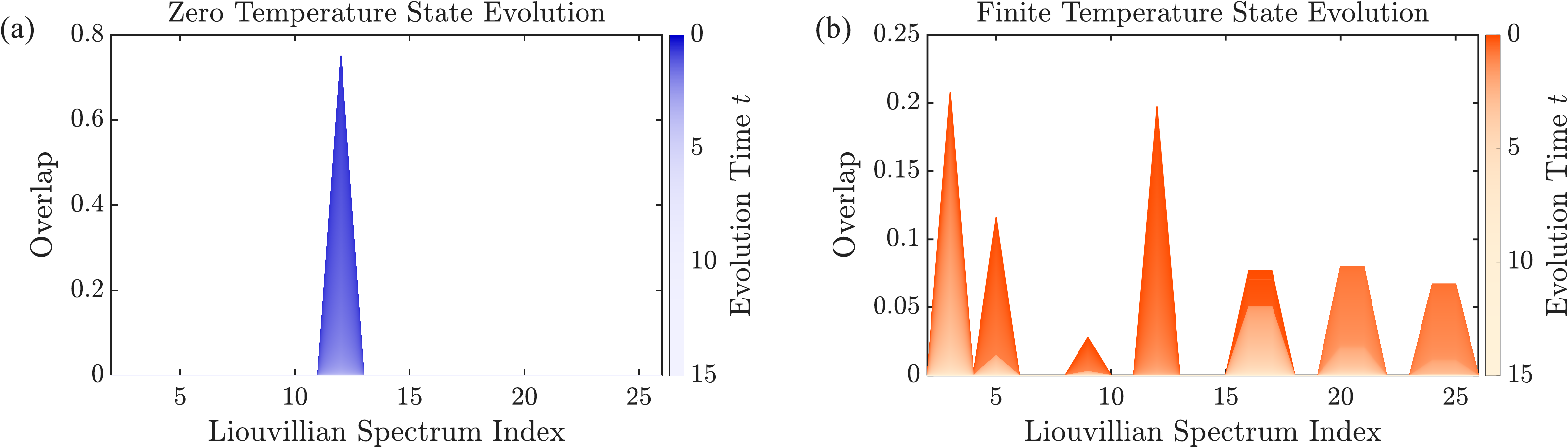}
	\caption{Time evolution of the spectral overlaps. The steady-state contribution is omitted and only transient modes \((k>1)\) are shown. Results are presented for the representative interaction range \(\alpha=4\). (a) Relaxation of the ground state \((T=0)\). The spectral weight is concentrated in the symmetry-selected fast-relaxation channel. (b) Relaxation of the finite-temperature state \((T=10)\). The state has finite overlap with several Liouvillian modes, including the slowest non-steady channel, which leads to a bottleneck in the relaxation dynamics.}
	\label{figA1}
\end{figure}

Figure~\ref{figA1} illustrates this mode-resolved decay. For the highly symmetric ground state, the spectral weight is confined to the symmetry-selected transient channels and rapidly flows toward the leading accessible mode. In contrast, the finite-temperature state has nonzero weight on several slow Liouvillian modes, including the slowest non-steady channel. This finite slow-mode overlap acts as a dynamical bottleneck and produces the slower relaxation observed in the main text.

\section*{Symmetry Analysis of the Operator $O^{(2q)}$ and the $\lambda=-2q$ Mode}
\label{App3}

To elucidate the origin of the universal $\lambda=-2q$ decay rates, we begin with the minimal case $q=1$ and reconstruct the operator $O^{(2)}$ from symmetry considerations. Rather than treating $O^{(2)}$ only as a discrete sum of spin pairs, it can be reformulated in terms of the global $SU(2)$ generators. Defining the total spin raising and lowering operators as $S^\pm = \sum_i \sigma_i^\pm$ and the total magnetization as $S^z = \frac{1}{2}\sum_i \sigma_i^z$, their product yields
\begin{equation}
	\label{Aeq:SplusSminus}
	S^+ S^- = \sum_i \sigma_i^+ \sigma_i^- + \sum_{i \neq j} \sigma_i^+ \sigma_j^- .
\end{equation}
Recognizing that $\sigma_i^+ \sigma_i^- = (I + \sigma_i^z)/2$, we identify the second term as the operator $O^{(2)}$. Thus,
\begin{equation}
	\label{Aeq:O2_collective}
	O^{(2)} = S^+ S^- - S^z - \frac{L}{2},
\end{equation}
where $L$ is the system size. This expression shows that $O^{(2)}$ is constructed entirely from total spin operators and therefore has a globally uniform structure. In particular, \(O^{(2)}\) is invariant not only under lattice translations but also under the full permutation group \(S_L\), since every lattice site enters with equal weight.

The Liouvillian evolution is partitioned into a coherent part and a dissipative part,
\begin{equation}
	\mathcal{L}(\rho)=-i[H,\rho]+\mathcal{L}_{\mathrm{diss}}(\rho).
\end{equation}
For the isotropic \((\Delta=1)\) XXZ chain, the Hamiltonian \(H\) has global \(SU(2)\) symmetry and hence commutes with all total spin generators,
\begin{equation}
	[H,S^\pm]=[H,S^z]=0 .
\end{equation}
Consequently, \(H\) also commutes with \(O^{(2)}\):
\begin{equation}
	\label{Aeq:H_O2_zero}
	[H, O^{(2)}] = [H, S^+ S^- - S^z - L/2] = 0 .
\end{equation}
This commutation relation implies that \(O^{(2)}\) is unaffected by the coherent Hamiltonian evolution. Therefore, regardless of the interaction range \(\alpha\), as long as the global \(SU(2)\) symmetry is preserved, the coherent contribution to the dynamics of \(O^{(2)}\) vanishes identically.

The dissipative contribution \(\mathcal{L}_{\mathrm{diss}}\) is determined by the local dephasing jump operators \(P_l = |\uparrow_l\rangle\langle\uparrow_l|\). The action of the Lindblad dissipator on an arbitrary operator \(X\) is
\begin{equation}
	\mathcal{L}_{\mathrm{diss}}(X)
	=
	\sum_l
	\left(
	2P_l X P_l-\{P_l,X\}
	\right).
\end{equation}
For a single-site operator \(\sigma_m^+\), a straightforward calculation gives
\begin{equation}
	\mathcal{L}_{\mathrm{diss}}(\sigma_m^+)=-\sigma_m^+,
\end{equation}
and similarly
\begin{equation}
	\mathcal{L}_{\mathrm{diss}}(\sigma_n^-)=-\sigma_n^- .
\end{equation}
For the two-body correlations appearing in \(O^{(2)}\), only the dephasing channels acting on sites \(m\) and \(n\) contribute. Therefore,
\begin{equation}
	\label{Aeq:diss_sigma_plus_minus}
	\begin{aligned}
		\mathcal{L}_{\mathrm{diss}}(\sigma_m^+ \sigma_n^-)
		&=
		[\mathcal{L}_{\mathrm{diss},m}(\sigma_m^+)]\sigma_n^-
		+
		\sigma_m^+[\mathcal{L}_{\mathrm{diss},n}(\sigma_n^-)]  \\
		&=
		-2\sigma_m^+\sigma_n^- .
	\end{aligned}
\end{equation}
Summing over all \(m\neq n\), we obtain
\begin{equation}
	\mathcal{L}_{\mathrm{diss}}(O^{(2)})=-2O^{(2)} .
\end{equation}
Combined with the vanishing commutator in Eq.~\eqref{Aeq:H_O2_zero}, the total Liouvillian action becomes
\begin{equation}
	\label{Aeq9}
	\mathcal{L}(O^{(2)}) = -2 O^{(2)} .
\end{equation}

This derivation explains three key features of the \(\lambda=-2\) mode. First, the decay rate is independent of the system size, because each two-site spin-flip coherence \(\sigma_m^+\sigma_n^-\) acquires the same dephasing-induced decay rate \(-2\). Second, the corresponding eigenoperator is globally uniform, which follows directly from the \(SU(2)\) algebraic structure. Third, the analysis clarifies the transition from degeneracy to an isolated symmetry-selected mode. In the all-to-all limit \((\alpha=0)\), the permutation symmetry \(S_L\) allows a broad family of operator structures of the form
\begin{equation}
	\sum_{i\neq j} C_{ij}\sigma_i^+\sigma_j^-
\end{equation}
to commute with \(H\), leading to a highly degenerate \(\lambda=-2\) subspace. However, for \(\alpha>0\), the spatial dependence of the interaction breaks the full permutation symmetry. Generic nonuniform combinations no longer commute with the Hamiltonian. The uniform combination \(O^{(2)}\), by contrast, remains compatible with the global \(SU(2)\) structure and survives as the isolated symmetry-selected fast-relaxation mode governing the high-symmetry relaxation channel. This analytical framework naturally generalizes to the family of higher-order operators \(O^{(2q)}\), which correspond to fast-relaxation channels with decay rates \(\lambda=-2q\).

\section*{Equivalence between $O^{(2q)}$ and the $\lambda=-2q$ Eigenmode}
\label{App4}

To establish that the Liouvillian eigenmode at \(\lambda=-2q\) is identical to the operator \(O^{(2q)}\), we analyze how coherent Hamiltonian evolution lifts the dissipative degeneracy. To make this mechanism explicit, we first focus on the correspondence between the Liouvillian eigenmode at \(\lambda=-2\) and the operator \(O^{(2)}\).

First, in the purely dissipative limit \((H=0)\), any operator of the form
\begin{equation}
	A = \sum_{i\neq j} C_{ij}\, \sigma_i^+ \sigma_j^-
\end{equation}
is an eigenvector of the dephasing dissipator \(\mathcal{L}_{\rm diss}\) with eigenvalue \(-2\). This yields a degenerate manifold of dimension \(L(L-1)\), within which the uniform operator \(O^{(2)}\), corresponding to \(C_{ij}=1\), is one specific state. This explains the highly degenerate \(\lambda=-2\) spectrum in the all-to-all limit \((\alpha=0)\).

For \(\alpha>0\), the long-range Hamiltonian
\begin{equation}
	H = -\sum_{m<n} J_{mn} \mathbf{S}_m \cdot \mathbf{S}_n, \qquad J_{mn}>0,
\end{equation}
induces coherent dynamics through the commutator part of the Liouvillian,
\begin{equation}
	\mathcal{L}_{\rm coh}(\cdot)=-i[H,\cdot],
\end{equation}
which lifts the dissipative degeneracy. A general operator \(A\) remains an exact \(\lambda=-2\) eigenmode only if
\begin{equation}
	[H,A]=0,
\end{equation}
namely if it is conserved under the Hamiltonian dynamics. Since the spatially dependent couplings \(J_{mn}\) break full permutation symmetry, a generic nonuniform coefficient pattern \(C_{ij}\) leads to a nonzero commutator and hybridization with other Liouvillian modes. The uniform choice \(C_{ij}=1\), however, recovers \(O^{(2)}\). Because the isotropic model \((\Delta=1)\) preserves global \(SU(2)\) symmetry, \(H\) commutes with all total-spin generators, ensuring
\begin{equation}
	[H,O^{(2)}]\equiv 0
\end{equation}
for arbitrary interaction range \(\alpha>0\). Thus, among the degenerate dissipative modes, \(O^{(2)}\) remains pinned at \(\lambda=-2\) as an isolated symmetry-selected eigenmode.

\begin{figure}[ht!]
	\centering
	\includegraphics[width=0.75\linewidth]{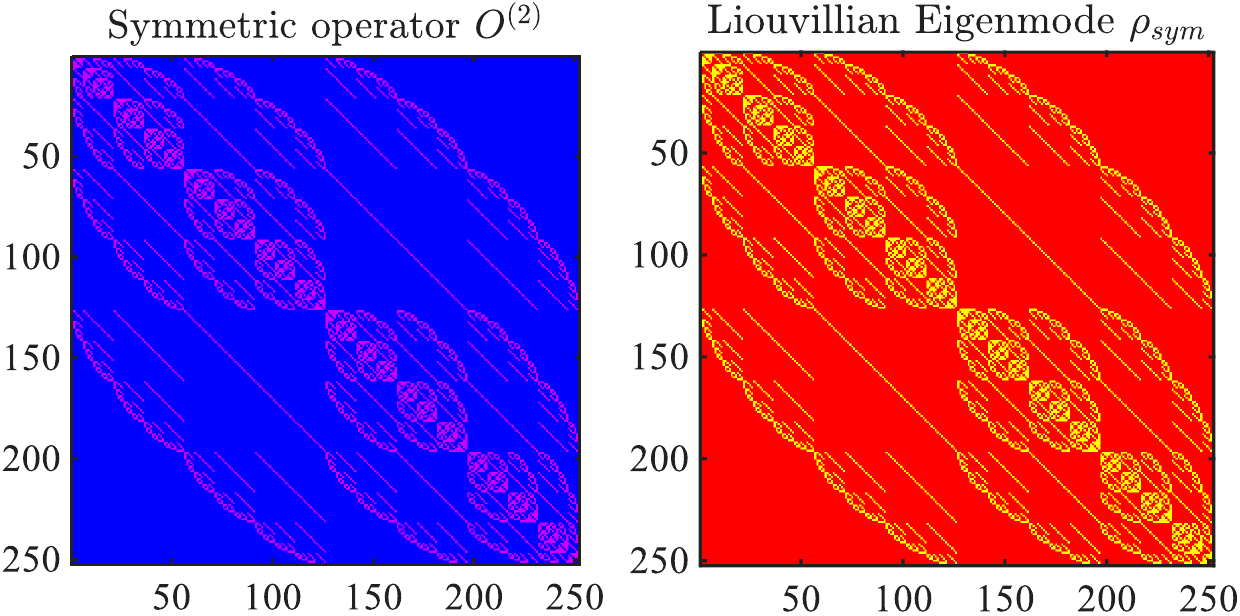}
	\caption{
		Matrix representation of the analytically constructed symmetric operator \(O^{(2)}\) (left) and the numerically obtained Liouvillian eigenmode at \(\lambda=-2\) (right).
		The identical support and block structure confirm that the universal fast-relaxation channel is governed by the symmetry-selected mode \(O^{(2)}\).
	}
	\label{figS1}
\end{figure}

We verify this correspondence numerically. Figure~\ref{figS1} compares the matrix representation of the analytically constructed \(O^{(2)}\) with the numerically obtained Liouvillian eigenmode at \(\lambda=-2\). Both operators exhibit the same support, block structure, and symmetry-sector organization. Using the Hilbert--Schmidt inner product
\begin{equation}
	\langle A,B\rangle=\mathrm{Tr}(A^\dagger B),
\end{equation}
we find the cosine similarity to be unity within numerical precision. This confirms that the analytical operator \(O^{(2)}\) and the numerical \(\lambda=-2\) eigenmode are collinear in Liouville space. The universal fast-relaxation channel is therefore governed by the symmetry-selected mode \(O^{(2)}\).

The same structural correspondence naturally generalizes to the higher-order sequence of operators \(O^{(2q)}\) and Liouvillian eigenvalues \(\lambda=-2q\). Here \(O^{(2q)}\) describes a \(2q\)-spin coherence whose dissipative decay rate is fixed by the number of spin-flip operators, while the Hamiltonian \(SU(2)\) symmetry keeps the uniform combination decoupled from coherent mixing.

\begin{figure*}
	\centering
	\includegraphics[width=1\linewidth]{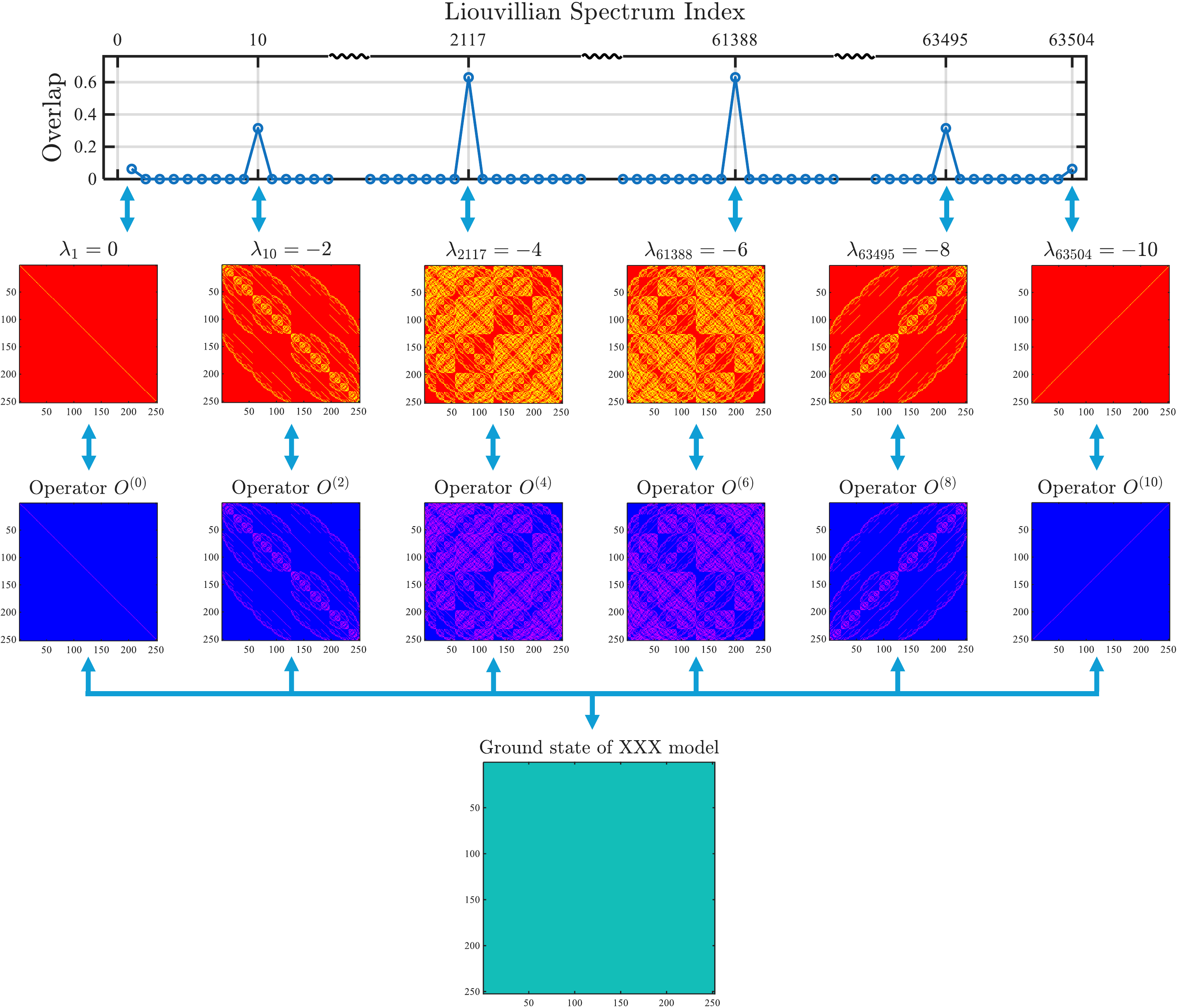}
	\caption{
		Symmetry-sector structure of the Liouvillian modes involved in the fast-relaxation dynamics. When an initially high-symmetry state evolves under local dephasing, the dynamics is restricted to the magnetization-conserving \(U(1)\) sectors. The relevant Liouvillian eigenmodes correspond to the higher-order operators \(O^{(2q)}\). A coherent linear combination of these zero-\(U(1)\)-charge operators reconstructs the initial high-symmetry configuration and provides the symmetry-selected relaxation channel. The calculation is performed by exact diagonalization for \(L=10\) in the zero-magnetization subspace \(S^z=0\).
	}
	\label{figA3}
\end{figure*}

To further illustrate how this symmetry-selected mechanism operates within the Liouvillian spectrum, we examine the symmetry-sector structure of the relevant modes, as shown in Fig.~\ref{figA3}. Although local dephasing reduces the manifest symmetry of the Liouvillian dynamics to the \(U(1)\) subgroup associated with magnetization conservation, the initial high-symmetry state can still be represented through a coherent combination of Liouvillian eigenmodes carrying the appropriate \(U(1)\) quantum numbers. These modes are naturally connected to the higher-order operators \(O^{(2q)}\), which organize the fast-relaxation channels selected by the underlying global symmetry. These operators lie in the zero-\(U(1)\)-charge sector, i.e., they satisfy \([S^z,O^{(2q)}]=0\).

\section*{Exact Arbitrary-\(q\) Left Eigenoperators}
\label{App:exact_q_modes}

We now make explicit that the operators \(O^{(2q)}\) form exact \emph{left} Liouvillian eigenoperators. This is the relevant statement for the spectral coefficients in Eq.~\eqref{Aeq3}, because the expansion coefficients are determined by the left eigenoperators \(\hat l_k\). We write the uniform \(2q\)-spin coherence operator in set notation as
\begin{equation}
	O^{(2q)}
	=
	\sum_{\substack{A,B\subset\{1,\ldots,L\}\\
	|A|=|B|=q,\ A\cap B=\emptyset}}
	\left(\prod_{i\in A}\sigma_i^+\right)
	\left(\prod_{j\in B}\sigma_j^-\right),
	\label{Aeq:O2q_set}
\end{equation}
up to an overall normalization. This expression is equivalent to the ordered-index definition used in the main text, differing only by a factor of \((q!)^2\).

At the ferromagnetic isotropic point, the Hamiltonian is
\begin{equation}
	H=-\sum_{i<j}J_{ij}\,\mathbf S_i\cdot\mathbf S_j,
	\qquad
	J_{ij}=\frac{J}{|i-j|^\alpha},
	\qquad J>0 .
	\label{Aeq:ferro_H_q}
\end{equation}
For arbitrary spatial dependence of \(J_{ij}\), the isotropic Hamiltonian has global \(SU(2)\) symmetry and therefore commutes with the collective generators,
\begin{equation}
	[H,S^\pm]=[H,S^z]=0 .
	\label{Aeq:H_collective_q}
\end{equation}
The operator \(O^{(2q)}\) is invariant under arbitrary permutations of lattice sites, since Eq.~\eqref{Aeq:O2q_set} sums over all disjoint subsets \(A\) and \(B\) with fixed cardinality. For a spin-\(1/2\) chain, the algebra of permutation-invariant operators is generated by the collective spin operators \(S^\pm\) and \(S^z\). Hence \(O^{(2q)}\) can be expressed as a polynomial in the collective spin algebra,
\begin{equation}
	O^{(2q)}=F_q(S^+,S^-,S^z),
	\label{Aeq:O2q_collective_poly}
\end{equation}
where the explicit form of \(F_q\) is not needed. Combining Eqs.~\eqref{Aeq:H_collective_q} and \eqref{Aeq:O2q_collective_poly} gives
\begin{equation}
	[H,O^{(2q)}]=0 .
	\label{Aeq:H_O2q_zero}
\end{equation}
Thus the coherent part of the adjoint Liouvillian does not mix the spatially uniform \(O^{(2q)}\) operator with nonuniform coherence patterns.

We next evaluate the local dephasing contribution. The jump operators are projectors \(P_l=(1+\sigma_l^z)/2\), and the adjoint dissipator acts as
\begin{equation}
	\mathcal D^\dagger(X)
	=
	\sum_l\left(2P_lXP_l-\{P_l,X\}\right).
	\label{Aeq:Ddagger_q}
\end{equation}
On a single transverse spin operator one has
\begin{equation}
	\mathcal D_l^\dagger(\sigma_l^\pm)=-\sigma_l^\pm,
	\qquad
	\mathcal D_m^\dagger(\sigma_l^\pm)=0\quad(m\neq l).
	\label{Aeq:single_site_decay_q}
\end{equation}
Each monomial in Eq.~\eqref{Aeq:O2q_set} contains exactly \(q\) raising and \(q\) lowering operators on \(2q\) distinct sites. Therefore
\begin{equation}
	\mathcal D^\dagger
	\left[
	\left(\prod_{i\in A}\sigma_i^+\right)
	\left(\prod_{j\in B}\sigma_j^-\right)
	\right]
	=
	-2q
	\left(\prod_{i\in A}\sigma_i^+\right)
	\left(\prod_{j\in B}\sigma_j^-\right).
	\label{Aeq:monomial_decay_q}
\end{equation}
After summing over all disjoint subsets, this gives
\begin{equation}
	\mathcal D^\dagger(O^{(2q)})=-2q\,O^{(2q)} .
	\label{Aeq:D_O2q_q}
\end{equation}
Using Eq.~\eqref{Aeq:H_O2q_zero}, the adjoint Liouvillian satisfies
\begin{equation}
	\mathcal L^\dagger(O^{(2q)})
	=
	i[H,O^{(2q)}]+\mathcal D^\dagger(O^{(2q)})
	=
	-2q\,O^{(2q)} .
	\label{Aeq:Ldagger_O2q_q}
\end{equation}
Therefore \(O^{(2q)}\) is an exact left Liouvillian eigenoperator with eigenvalue \(\lambda_q=-2q\). This result is independent of the system size and of the interaction range \(\alpha\), provided the Hamiltonian remains isotropic and globally \(SU(2)\)-symmetric.

\section*{Analytical Derivation of the Overlap Comb Distribution}
\label{App:overlap_comb}

We now derive the exact overlap between the ferromagnetic ground state at the isotropic point and the \(\lambda=-2q\) left Liouvillian eigenoperators. Since the ferromagnetic Hamiltonian in Eq.~\eqref{Aeq:ferro_H_q} has a maximum-spin ground-state multiplet, we work in a fixed magnetization sector. In the zero-magnetization sector, the symmetry-unbroken ground-state representative is the Dicke state
\begin{equation}
	|D_L^{(0)}\rangle
	=
	\left|S=\frac{L}{2},M=0\right\rangle
	=
	\frac{1}{\sqrt{\mathcal D_L}}
	\sum_{\alpha\in S^z=0}|\alpha\rangle,
	\qquad
	\mathcal D_L=\binom{L}{L/2} .
	\label{Aeq:Dicke_M0}
\end{equation}
Its density matrix is therefore
\begin{equation}
	\rho_D
	=
	|D_L^{(0)}\rangle\langle D_L^{(0)}|
	=
	\frac{1}{\mathcal D_L}
	\sum_{\alpha,\beta\in S^z=0}|\alpha\rangle\langle\beta| .
	\label{Aeq:Dicke_density}
\end{equation}
The operator \(O^{(2q)}\) connects two zero-magnetization configurations that differ by exchanging \(q\) up spins and \(q\) down spins, equivalently by a Hamming distance \(2q\). For a fixed reference configuration, the number of such connected configurations is
\begin{equation}
	N_q=\binom{L/2}{q}^2 .
	\label{Aeq:Nq_M0}
\end{equation}
Let \(\widetilde O^{(2q)}\) denote the Hilbert--Schmidt normalized \(2q\)-coherence shell, normalized such that \(\mathrm{Tr}[(\widetilde O^{(2q)})^\dagger\widetilde O^{(2q)}]=1\). The left-mode overlap amplitude is then
\begin{equation}
	c_q
	=
	\mathrm{Tr}\left[(\widetilde O^{(2q)})^\dagger\rho_D\right]
	=
	\frac{\binom{L/2}{q}}{\sqrt{\binom{L}{L/2}}},
	\qquad
	q=0,1,\ldots,\frac{L}{2} .
	\label{Aeq:cq_exact}
\end{equation}
The normalized spectral weights are
\begin{equation}
	w_q=|c_q|^2
	=
	\frac{\binom{L/2}{q}^2}{\binom{L}{L/2}},
	\qquad
	\sum_{q=0}^{L/2}w_q=1,
	\label{Aeq:wq_exact}
\end{equation}
where the normalization follows from Vandermonde's identity. The \(q=0\) component corresponds to the diagonal steady-state sector, while the transient relaxation comb starts from \(q=1\), with \(\lambda=-2\).

For the numerical case \(L=10\), Eq.~\eqref{Aeq:cq_exact} gives
\begin{equation}
	(c_0,c_1,c_2,c_3,c_4,c_5)
	=
	\frac{1}{\sqrt{252}}(1,5,10,10,5,1),
	\label{Aeq:L10_overlap}
\end{equation}
which explains the peak heights observed in the mode-overlap data.

The same counting argument also applies in a general fixed
magnetization sector \(M\). For \(S=L/2\), the Dicke state
\(|S,M\rangle\) contains
\[
N_\uparrow=S+M,\qquad N_\downarrow=S-M
\]
up and down spins, respectively, and the Hilbert-space
dimension in this fixed-magnetization sector is
\[
\mathcal D_M=\binom{2S}{S+M}.
\]
The \(2q\)-spin coherence shell connects two configurations
that differ by exchanging \(q\) up spins and \(q\) down spins.
For each reference configuration, the number of such connected
configurations is
\[
N_q(M)
=
\binom{S+M}{q}
\binom{S-M}{q}.
\]
After Hilbert--Schmidt normalization of each coherence shell,
the left-mode overlap amplitude is therefore
\begin{equation}
	c_q(M)
	=
	\sqrt{
	\frac{
	\binom{S+M}{q}\binom{S-M}{q}
	}{
	\binom{2S}{S+M}
	}
	},
	\qquad
	q=0,1,\ldots,S-|M| .
	\label{Aeq:cq_general_M}
\end{equation}
The corresponding normalized spectral weights are
\begin{equation}
	w_q(M)=|c_q(M)|^2
	=
	\frac{
	\binom{S+M}{q}\binom{S-M}{q}
	}{
	\binom{2S}{S+M}
	},
	\qquad
	\sum_{q=0}^{S-|M|}w_q(M)=1 .
	\label{Aeq:wq_general_M}
\end{equation}
The normalization follows from the Vandermonde identity
\[
\sum_{q=0}^{S-|M|}
\binom{S+M}{q}
\binom{S-M}{q}
=
\binom{2S}{S+M}.
\]
For \(M=0\), Eq.~\eqref{Aeq:cq_general_M} reduces to
\[
c_q=
\frac{\binom{S}{q}}{\sqrt{\binom{2S}{S}}},
\]
which is the zero-magnetization result in Eq.~\eqref{Aeq:cq_exact}.

This expression reduces to Eq.~\eqref{Aeq:wq_exact} for \(M=0\), and shows that the symmetry-selected overlap structure is not restricted to the zero-magnetization sector.

As a check away from zero magnetization, consider
\(L=10\), \(N_\uparrow=3\), and \(N_\downarrow=7\), corresponding
to \(S=5\) and \(M=-2\). Equation~\eqref{Aeq:cq_general_M}
gives
\begin{equation}
	(c_0,c_1,c_2,c_3)
	=
	\left(
	\sqrt{\frac{1}{120}},
	\sqrt{\frac{21}{120}},
	\sqrt{\frac{63}{120}},
	\sqrt{\frac{35}{120}}
	\right)
	\label{Aeq:L10_Mminus2_overlap}
\end{equation}
in agreement with the numerical left-eigenmode overlaps.

The thermodynamic limit follows from a large-\(S\) saddle-point expansion. For \(L=2S\) and \(x=q/S\), Stirling's formula gives
\begin{equation}
\begin{split}
    w_q &\simeq
    \frac{1}{2\sqrt{\pi S}\,x(1-x)}
    \exp\left[-2S\bigl(\ln2-H(x)\bigr)\right],
    \\
    H(x)&=-x\ln x-(1-x)\ln(1-x).
\end{split}
\label{Aeq:wq_large_deviation}
\end{equation}
The distribution is peaked at \(x=1/2\), i.e., \(q=S/2=L/4\). Expanding near the maximum gives the Gaussian form
\begin{equation}
	w_q
	\simeq
	\frac{2}{\sqrt{\pi S}}
	\exp\left[-\frac{4(q-S/2)^2}{S}\right].
	\label{Aeq:wq_gaussian_S}
\end{equation}
Equivalently,
\begin{equation}
	w_q
	\simeq
	\frac{1}{\sqrt{2\pi\sigma_q^2}}
	\exp\left[-\frac{(q-L/4)^2}{2\sigma_q^2}\right],
	\qquad
	\sigma_q^2\simeq\frac{L}{16}.
	\label{Aeq:wq_gaussian_L}
\end{equation}
The exact finite-size variance is
\begin{equation}
	\mathrm{Var}(q)=\frac{S^2}{4(2S-1)}=\frac{L^2}{16(L-1)}.
	\label{Aeq:variance_exact}
\end{equation}
Thus the symmetry-selected relaxation comb has an analytically determined Gaussian envelope in the large-system limit. Although the initial spectral weight is concentrated around \(q\simeq L/4\), these components decay as \(e^{-2qt}\). Consequently, the long-time trace-distance exponent is governed by the slowest accessible transient tooth, \(q=1\), corresponding to the universal decay rate \(\lambda=-2\). The trace norm itself is nonlinear, so the formulas above should be understood as the linear Liouville-space spectral weights; they determine the asymptotic exponent through the slowest nonzero accessible component.

\begin{figure}
	\centering
	\includegraphics[width=0.98\linewidth]{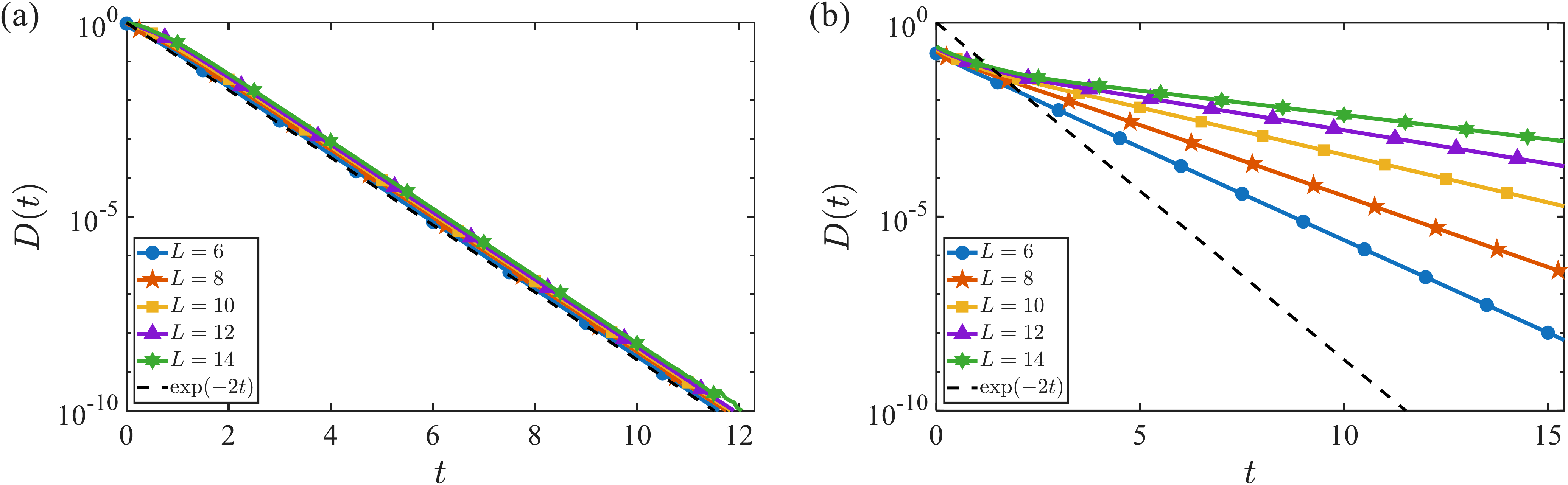}
	\caption{
		Finite-size scaling of relaxation dynamics. The time evolution of the trace distance \(D(t)\) is compared for different lattice sizes \(L=6,8,10,12,14\) at a representative interaction exponent \(\alpha=4\).
		(a) Relaxation from the highly symmetric ground state \((T=0)\). The dynamical trajectories collapse onto the universal \(e^{-2t}\) decay curve, showing that the dominant symmetry-selected relaxation channel is insensitive to system size.
		(b) Relaxation from the finite-temperature state \((T=10)\). In contrast, the relaxation dynamics exhibits a clear dependence on system size, reflecting contributions from additional Liouvillian modes that are not fully constrained by the same symmetry-selection mechanism.
	}
	\label{figA4}
\end{figure}

\section*{Degeneracy Analysis in the All-to-All Limit ($\alpha=0$)}
\label{App5}

For any finite $\alpha>0$, the system is primarily characterized by spatial structure and global $SU(2)$ symmetry. However, in the $\alpha=0$ limit, the interaction becomes distance independent, transforming the model into an all-to-all isotropic XXX chain, i.e., the isotropic limit of the Lipkin--Meshkov--Glick (LMG) model. In this regime, the Hamiltonian can be written in terms of the square of the total spin:
\begin{equation}
	\label{Aeq11}
	\begin{aligned}
		H(\alpha=0)
		&= -\frac{J}{L}\sum_{i<j}\mathbf{S}_i\cdot\mathbf{S}_j \\
		&= -\frac{J}{2L}\left(\mathbf S_{\mathrm{tot}}^2-\frac{3L}{4}\right),\qquad J>0 .
	\end{aligned}
\end{equation}
This Hamiltonian possesses an enlarged permutation symmetry \(S_L\): it is invariant under the exchange of any two lattice sites.

As discussed above, the dephasing dissipator fixes the decay rate of any operator containing two transverse spin components, such as \(\sigma_i^+\sigma_j^-\), to exactly \(-2\). For \(\alpha>0\), the spatial dependence of the interaction \(J_{ij}\) generally removes the degeneracy among different spatial structures. In this case, the fully symmetric operator
\begin{equation}
	O^{(2)}=\sum_{i\neq j}\sigma_i^+\sigma_j^-
\end{equation}
is singled out by the global \(SU(2)\) structure and remains pinned at \(\lambda=-2\). In contrast, when \(\alpha=0\), the Hamiltonian treats all sites identically. The enlarged permutation symmetry allows many additional operator combinations within the two-spin coherence sector to commute with \(H\). Consequently, the coherent part of the Liouvillian no longer distinguishes these spatial structures, and many modes that are split for \(\alpha>0\) collapse into a highly degenerate manifold at the dissipative eigenvalue \(\lambda=-2\).

This symmetry also constrains the structure of the initial states. At \(\alpha=0\), the energy eigenvalues \(E_n\) depend only on the total spin quantum number \(S\). The ground state \(\rho_0\) belongs to a highly symmetric total-spin sector, while the finite-temperature state \(\rho_T\) is a statistical mixture over different \(S\) sectors. Because the Boltzmann weights \(e^{-\beta E_n}\) depend only on \(S\), the thermal state preserves the same global permutation symmetry of the all-to-all Hamiltonian.

The spectral overlap
\begin{equation}
	c_k=\mathrm{Tr}(l_k^\dagger\rho)
\end{equation}
is nonzero only for Liouvillian modes \(l_k\) whose symmetry is compatible with that of the initial state. At \(\alpha=0\), \(\rho_0\) and \(\rho_T\) have the same permutation-symmetric character in Liouville space. Increasing the temperature redistributes statistical weight among different energy sectors, but it does not introduce spatially nonuniform structures. Consequently, the two states predominantly excite the same symmetry-allowed decay manifold, so the spectral separation that distinguishes their dominant relaxation channels for \(\alpha>0\) is lost, as shown in the inset of Fig.~\ref{fig3}.

The physical origin of the strong quantum Mpemba effect (SQME) can therefore be understood as a competition between thermal fluctuations and spatial dispersion. For \(\alpha>0\), the distance dependence of the interactions provides the spatial structure needed to differentiate the relaxation of thermal states from that of the ground state. Finite-temperature states can acquire projections onto nonuniform Liouvillian modes, modifying their overlap with the symmetry-selected \(\lambda=-2\) channel and allowing the ground state to overtake the thermal state. At \(\alpha=0\), this spatial dispersion is absent. The dominant relaxation channel becomes a highly degenerate permutation-symmetric manifold. Since \(\rho_0\) and \(\rho_T\) are not distinguished by spatial structure under the all-to-all symmetry, the dominant asymptotic decay channels become degenerate. Their long-time relaxation rates therefore collapse to the same \(e^{-2t}\) behavior, leading to the disappearance of the strong Mpemba separation in the all-to-all limit. These results suggest that anomalous relaxation in open many-body systems is driven by the interplay between thermal excitations, spatial structure, and Liouvillian spectral selection.

\begin{figure}
	\centering
	\includegraphics[width=0.6\linewidth]{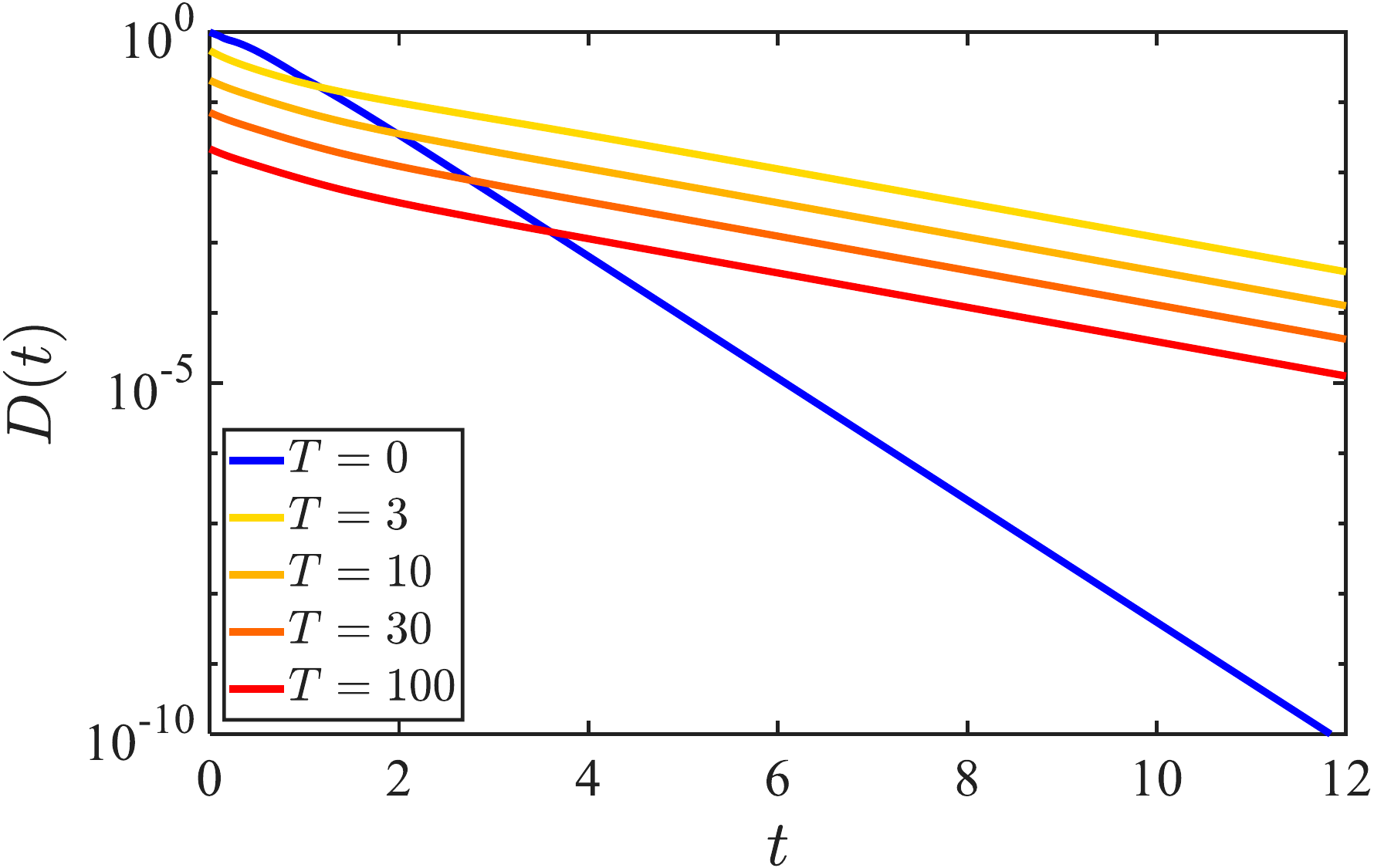}
	\caption{
		Strong quantum Mpemba effect. The relaxation dynamics are calculated for a long-range interaction exponent $\alpha=4$ and system size $L=10$, comparing the zero-temperature ground state ($T=0$) with finite-temperature initial states at $T=3, 10, 30,$ and $100$. The ground state consistently relaxes faster than the thermal states, demonstrating the strong quantum Mpemba effect for the temperature range considered.
	}
	\label{figA5}
\end{figure}

\section*{Finite-Size Scaling and Universality}
\label{App6}

Driven by the underlying symmetry-selection mechanism, the relaxation trajectory of the ground state is locked onto the universal \(e^{-2t}\) decay law and exhibits remarkable robustness against finite-size effects. To verify this scale robustness, we perform numerical simulations for different lattice sizes, \(L=6,8,10,12\), and \(14\).

As illustrated in Fig.~\ref{figA4}, the dynamical trajectories of the highly symmetric ground state remain nearly invariant across different system sizes. The small discrepancies observed at very early times originate from size-dependent variations in the initial projection weights onto higher-order symmetric operators \(O^{(2q)}\). However, because the Liouvillian eigenmodes associated with these higher-order channels have increasingly large decay rates,
\[
\mathrm{Re}(\lambda)=-4,-6,\ldots ,
\]
their contributions decay rapidly and become negligible in the asymptotic dynamics. As a result, the long-time relaxation is dominated by the universal \(\lambda=-2\) channel, producing the size-independent \(e^{-2t}\) decay.

In sharp contrast, the finite-temperature states \((T=10)\) display a distinct behavior. Because they are not fully restricted to the symmetry-selected fast-relaxation channel, their relaxation pathways exhibit a clear dependence on the lattice size and shift systematically with \(L\). This dichotomy reinforces the central picture of the main text: while thermal relaxation is sensitive to additional Liouvillian modes and finite-size spectral structure, the ground-state dynamics is governed by a robust symmetry-selected channel that remains essentially scale invariant.

\section*{Numerical Verification of the Strong Quantum Mpemba Effect}
\label{App7}

In the present comparison, the strong quantum Mpemba effect is identified by the fact that the initially farther zero-temperature ground state relaxes faster than the finite-temperature states considered here and overtakes them during the relaxation. We compare the relaxation dynamics of the zero-temperature ground state with those of finite-temperature states as they evolve toward the infinite-temperature steady state.

As illustrated in Fig.~\ref{figA5}, we compute the relaxation dynamics for a long-range interaction exponent $\alpha=4$, comparing the zero-temperature ground state against finite-temperature initial states at $T=3, 10, 30,$ and $100$. The finite-temperature states follow substantially slower relaxation trajectories, while the ground state undergoes accelerated relaxation and reaches the steady state faster within the time window shown. This trajectory crossing confirms the occurrence of the strong quantum Mpemba effect for the temperature range considered.

\begin{figure}
    \centering
    \includegraphics[width=0.98\linewidth]{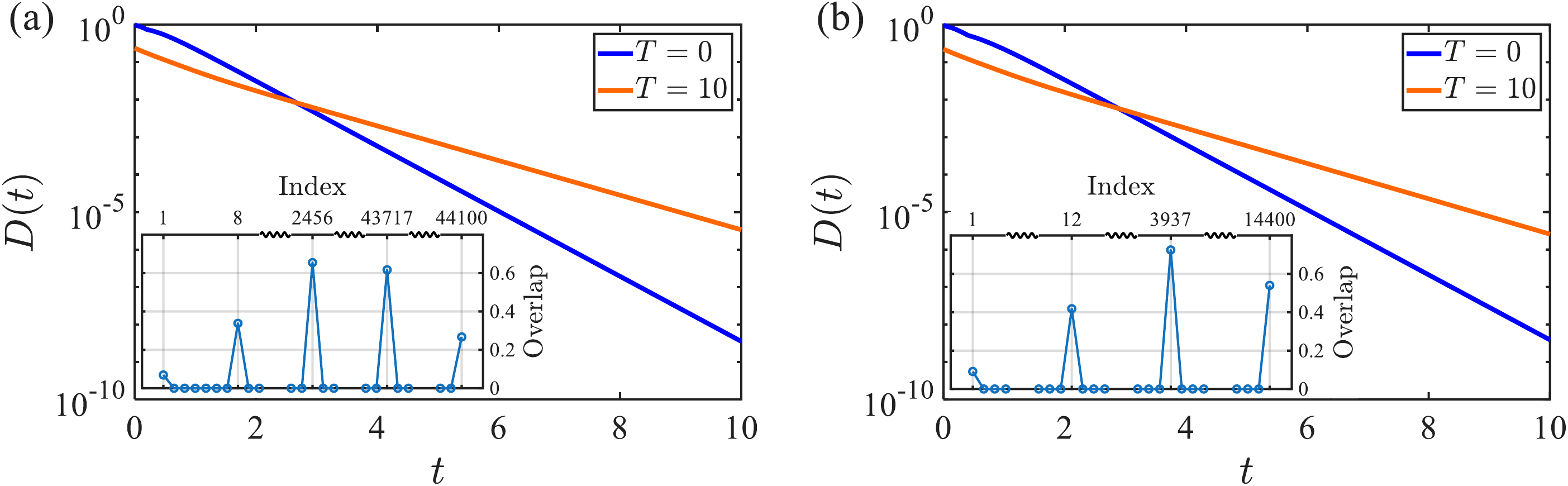}
    \caption{We find that the strong quantum Mpemba effect persists in sectors with nonzero total magnetization ($L=10$). Panels (a) and (b) correspond to the cases of total magnetizations $M=-1$ and $M=-2$, respectively. The insets show the overlap between the ground-state density matrix and the Liouvillian eigenmodes. The overlap coefficients follow the exact Fock-moment formula, Eq.~\eqref{Aeq:wq_general_M}.}
    \label{figA6}
\end{figure}


\begin{table*}[t]
\caption{
Overlap between the initial state and the Liouvillian eigenmodes
$O^{(2q)}$ in different magnetization sectors $|M|$. Here $M$ is the eigenvalue of $S^z$ with $S^z = \frac{1}{2} \sum_i \sigma_i^z$. For $L = 10$, the rows $|M| = 0, 1, 2, 3, 4$ correspond to $\{N_1, N_2\} = \{5, 5\}, \{4, 6\}, \{3, 7\}, \{2, 8\}, \{1, 9\}$. A dash indicates the absence of the corresponding mode, while ``deg.'' denotes a degenerate eigensubspace.
}
\label{tab:overlap_spin_sectors}
\centering
\begin{ruledtabular}
\begin{tabular}{c@{\hspace{1.5em}}cccccc}
$|M|$
& $O^{(0)}$
& $O^{(2)}$
& $O^{(4)}$
& $O^{(6)}$
& $O^{(8)}$
& $O^{(10)}$
\\
\hline
$0$
& $0.0629941$
& $0.314970$
& $0.629941$
& $0.629941$
& $0.314970$
& $0.0629941$
\\
$1$
& $0.0690060$
& $0.338062$
& $0.654654$
& $0.617213$
& $0.267261$
& \multicolumn{1}{c}{---}
\\
$2$
& $0.0912871$
& $0.418330$
& $0.724569$
& $0.540062$
& \multicolumn{1}{c}{---}
& \multicolumn{1}{c}{---}
\\
$3$
& $0.149071$
& $0.596285$
& $0.788811$
& \multicolumn{1}{c}{---}
& \multicolumn{1}{c}{---}
& \multicolumn{1}{c}{---}
\\
$4$
& $0.316228$
& \multicolumn{1}{c}{$\mathrm{deg.}$}
& \multicolumn{1}{c}{---}
& \multicolumn{1}{c}{---}
& \multicolumn{1}{c}{---}
& \multicolumn{1}{c}{---}
\end{tabular}
\end{ruledtabular}
\end{table*}

\section*{Structure of the Ground-State Density Matrix at Nonzero Magnetization}
\label{App8}

The sQME persists even at nonzero total magnetization. In a fixed magnetization sector, the symmetry-unbroken ground-state representative is the Dicke state $|S,M\rangle$, with $N_\uparrow=S+M$ and $N_\downarrow=S-M$. Its density matrix has nonzero projection only onto the normalized coherence shells $\widetilde O^{(2q)}$ with
\[
q=0,1,\ldots,S-|M|.
\]
Equivalently, within the symmetry-selected comb subspace,
\[
|S,M\rangle\langle S,M|
=
\sum_{q=0}^{S-|M|}c_q(M)\widetilde O^{(2q)},
\]
where the coefficients are given by Eq.~\eqref{Aeq:cq_general_M}. Thus the accessible comb is truncated at the eigenvalues
\[
\lambda=0,-2,\ldots,-2(S-|M|).
\]
As representative examples, we consider a spin chain of length $L=10~(S=5)$ with total magnetizations $M=-1~(N_\uparrow=4)$ and $M=-2~(N_\uparrow=3)$, shown in Fig.~\ref{figA6}. For $M=-1$, the ground-state density matrix has finite overlap only with the Liouvillian eigenmodes whose eigenvalues are $0,-2,-4,-6$, and $-8$. When the total magnetization is reduced to $M=-2$, the overlap is further restricted to eigenmodes with eigenvalues $0,-2,-4$, and $-6$. These eigenmodes correspond to the allowed $O^{(2q)}$ coherence shells within the fixed-magnetization sector.


\begin{figure}[h!]
    \centering
    \includegraphics[width=0.98\linewidth]{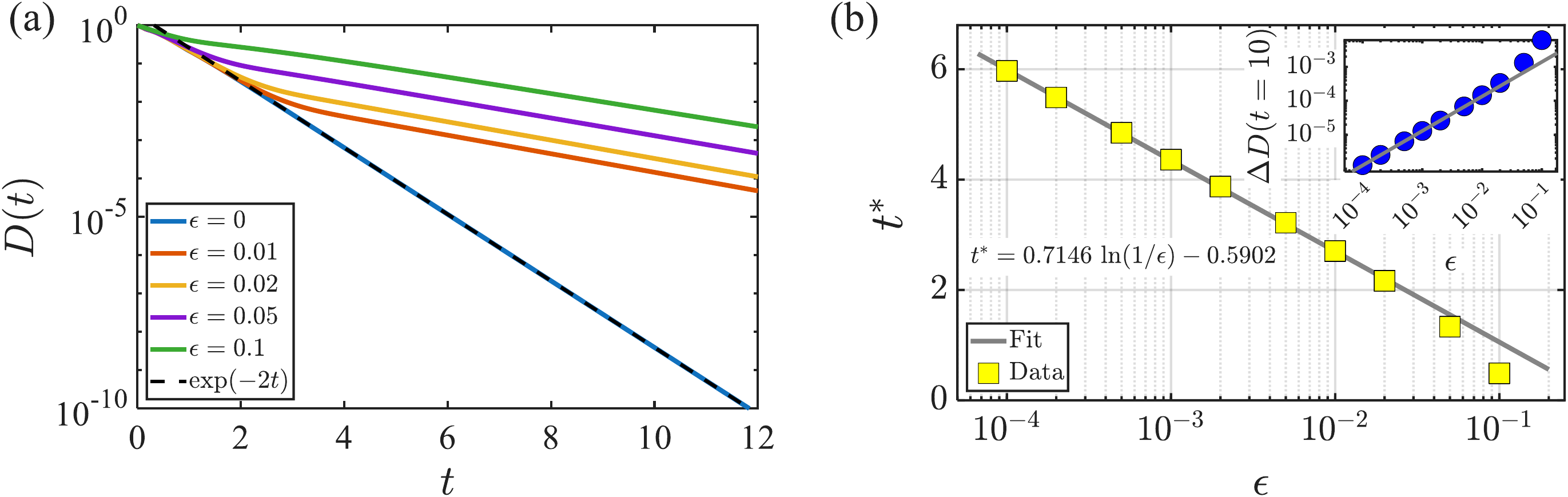}
    \caption{Robustness of the comb dynamics against weak SU(2) symmetry breaking. (a) Trace distance $D(t)$ for $\Delta=1+\epsilon$ ($L=10$, $\alpha=4$, initial Dicke state). The black curve ($\epsilon=0$) follows the closed-form comb solution $D_0(t)\propto e^{-2t}$, while finite \(\epsilon\) activates a slow relaxation tail and leaves the early-time comb decay nearly unchanged in the weak-breaking regime. The crossover time $t^{*}$ separates the comb-dominated regime from the slow-tail regime.
(b) The extracted crossover time follows
\(t^\ast \simeq 0.7146\ln(1/\epsilon)-0.5902\), demonstrating that the
comb-dominated window decreases only logarithmically with the
symmetry-breaking strength. The largest-$\epsilon$ point is shown as a reference and reflects the onset of higher-order corrections beyond the leading weak-breaking scaling. \textit{Inset:} The deviation
$\Delta D(t_{0},\epsilon)=|D_{\epsilon}(t_{0})-D_{0}(t_{0})|$
scales as
$\Delta D\propto\epsilon^{1.03}$,
which is consistent with a leading linear activation of the slow Liouvillian mode, while its decay rate remains nearly unchanged
($\Gamma_{s}\simeq0.131$).}
    \label{fig:robust_comb}
\end{figure}

\section*{Robustness of the comb dynamics against weak SU(2) symmetry breaking}
\label{App9}


We further examine whether the comb relaxation is stable against weak
SU(2) symmetry breaking.  To this end, we perturb the SU(2)-symmetric
point by setting
\[
\Delta=1+\epsilon ,
\]
where \(\epsilon\) measures the strength of the symmetry-breaking
perturbation.  At \(\epsilon=0\), the Dicke-state dynamics is governed by
the exact comb solution and the trace distance follows
\[
D_0(t)\propto e^{-2t}.
\]
As shown in Fig.~\ref{fig:robust_comb}(a), the exact Liouvillian evolution
at the SU(2) point is indistinguishable from the closed-form comb result.

For finite but small \(\epsilon\), the initial decay remains locked to the
comb rate \(2\).  The main effect of weak SU(2) breaking is not to
immediately destroy the fast comb relaxation, but rather to activate a
slow relaxation tail at later times.  This behavior can be captured by the
phenomenological decomposition
\begin{equation}
D_\epsilon(t)
\simeq
A_2 e^{-2t}
+
A_s(\epsilon)e^{-\Gamma_s t},
\label{eq:two_component_robustness}
\end{equation}
where \(A_2 e^{-2t}\) denotes the original comb contribution, while
\(A_s(\epsilon)e^{-\Gamma_s t}\) is the slow-tail contribution activated by
SU(2) symmetry breaking.

The crossover time \(t^\ast\) is defined as the time at which the fitted
slow-tail contribution becomes comparable to the comb contribution,
\begin{equation}
A_2 e^{-2t^\ast}
=
A_s(\epsilon)e^{-\Gamma_s t^\ast}.
\label{eq:tstar_definition}
\end{equation}
This gives
\begin{equation}
t^\ast
=
\frac{\ln|A_2/A_s(\epsilon)|}{2-\Gamma_s}.
\label{eq:tstar_expression}
\end{equation}
If the activated slow-tail amplitude scales as
$A_s(\epsilon)\propto \epsilon^p$, then Eq.~\eqref{eq:tstar_expression} implies
\begin{equation}
t^\ast
\simeq
\frac{p}{2-\Gamma_s}\ln(1/\epsilon)
+
\mathrm{const}.
\label{eq:tstar_log_scaling}
\end{equation}
Therefore, weak symmetry breaking is expected to shorten the comb-dominated time window only logarithmically in \(\epsilon\).

This expectation is confirmed by the numerical data in
Fig.~\ref{fig:robust_comb}(b).  The extracted crossover time is well fitted by
\begin{equation}
t^\ast
\simeq
0.7146\ln(1/\epsilon)-0.5902 ,
\label{eq:tstar_fit}
\end{equation}
showing that the lifetime of the comb-dominated regime decreases only
logarithmically with the SU(2)-breaking strength.  The slight deviation at
larger \(\epsilon\) reflects the expected onset of higher-order corrections
beyond the leading weak-breaking regime.

We also quantify the activation of the slow tail through the fixed-time
deviation
\begin{equation}
\Delta D(t_0,\epsilon)
=
\left|D_\epsilon(t_0)-D_0(t_0)\right| .
\label{eq:deltaD_definition}
\end{equation}
For a fixed \(t_0\), Eq.~\eqref{eq:two_component_robustness} gives
\begin{equation}
\Delta D(t_0,\epsilon)
\propto
A_s(\epsilon)e^{-\Gamma_s t_0}
\propto
\epsilon^p .
\label{eq:deltaD_scaling}
\end{equation}
The inset of Fig.~\ref{fig:robust_comb}(b) shows that
$\Delta D(t_0=10,\epsilon)\propto \epsilon^{1.03}$, which is consistent with a nearly linear activation of the slow relaxation channel.  Meanwhile, the fitted slow decay rate remains nearly unchanged, $\Gamma_s\simeq 0.131$, indicating that weak SU(2) breaking mainly populates an existing slow Liouvillian mode rather than creating a new slow decay rate.

These results demonstrate that the comb dynamics is not a fragile
fine-tuned feature destroyed by infinitesimal SU(2) breaking.  Instead,
weak symmetry breaking introduces only a parametrically small slow-tail
amplitude, so the \(e^{-2t}\) comb relaxation survives over a broad
pre-asymptotic time window whose duration decreases only logarithmically
with \(\epsilon\).

\end{document}